\documentclass[aps,prd,twocolumn,showpacs,nofootinbib,superscriptaddress]{revtex4}
\usepackage{amssymb}
\usepackage{graphicx}

\newcommand{\md}{{\mathrm{d}}}

\begin{document}

\title{High-order quantum back-reaction and quantum cosmology with a positive cosmological constant}


\author{Martin Bojowald}
\affiliation{Institute for Gravitation and the Cosmos,
The Pennsylvania State University,
104 Davey Lab, University Park, PA 16802, USA}
\author{David Brizuela}
\affiliation{Institute for Gravitation and the Cosmos,
The Pennsylvania State University,
104 Davey Lab, University Park, PA 16802, USA}
\author{Hector H.~Hern\'andez}
\affiliation{Universidad Aut\'onoma de Chihuahua,
Facultad de Ingenier\'ia,
Nuevo Campus Universitario, Chihuahua 31125, M\'exico}
\author{Michael J.~Koop}
\affiliation{Institute for Gravitation and the Cosmos,
The Pennsylvania State University,
104 Davey Lab, University Park, PA 16802, USA}
\author{Hugo A.~Morales-T\'ecotl}
\affiliation{Departamento de F\'{\i}sica,
Universidad Aut\'onoma Metropolitana Iztapalapa,
San Rafael Atlixco 186, CP 09340,
M\'exico D.F., M\'exico}

\begin{abstract}
When quantum back-reaction by fluctuations, correlations and higher
moments of a state becomes strong, semiclassical quantum mechanics
resembles a dynamical system with a high-dimensional phase
space. Here, systematic computational methods to derive the dynamical
equations including all quantum corrections to high order in the
moments are introduced, together with a (deparameterized)
quantum cosmological example to illustrate some implications. The
results show, for instance, that the Gaussian form of an initial state
is maintained only briefly, but that the evolving state settles down
to a new characteristic shape afterwards. Remarkably,
even in the regime of large high-order moments,
we observe a strong convergence within all considered orders that
supports the use of this effective approach.
\end{abstract}

\maketitle

\section{Introduction}

Semiclassical approximations are of importance throughout physics in
order to extract new effects or potentially observable phenomena in
regimes in which quantum features are relevant but not dominant. All
aspects of quantum gravity currently considered for potential tests
fall in this class of situations. While deep quantum phases are
crucial for several conceptual problems, the current developments
remain so diverse that reliable conclusions are difficult to
draw. Semiclassical physics, on the other hand, is already useful for
this class of theories in order to demonstrate their consistency with
large-scale properties of the universe and to approach potential
low-energy observations for instance via details of the cosmic
microwave background. For this reason, the development of systematic
semiclassical approximations is one of the major requirements in
current quantum-gravity research, especially for approaches such as
loop quantum gravity whose underlying principles do not make use of a
continuum geometry and perturbations around it.

In loop quantum gravity and cosmology, the main tool available at a
dynamical level is that of effective equations which describe the
evolution of expectation values in a dynamical state \cite{EffAc}. As
in most quantum systems, especially interacting ones, quantum
fluctuations, correlations and higher moments of the state then
back-react on the evolution of expectation values, described in
effective equations by coupling terms between classical and quantum
degrees of freedom. With an infinite number of independent moments of
a state, there are infinitely many quantum degrees of freedom, making
the general system of equations difficult to analyze. In semiclassical
regimes, however, the infinite set of equations reduces to finite sets
at any given order of the expansion by powers of $\hbar$. In this way,
a systematic expansion arises that goes well beyond other examples
such as the WKB approximation. To first order in $\hbar$ and combined
with a second-order adiabatic approximation, applied to anharmonic
oscillators, equations equivalent to those of the low-energy effective
action are produced when expanding around the ground state
\cite{EffAc,EffAcQM}. (The more widely applied WKB approximation does
not reproduce the low-energy effective action \cite{EffAcWKB}.) For
other systems and different states, on the other hand, the canonical
methods developed in \cite{EffAc} are more general and apply to
quantum cosmology as well, even taking into account the totally
constrained nature of these systems \cite{EffCons}. One should thus
expect that an application to quantum gravity allows comparisons with
low-energy effective quantum gravity
\cite{EffectiveNewton,EffectiveGR}, but can also highlight new effects
related to specific properties of quantum space-time structure as
opposed to just graviton dynamics on a classical background.

Compared with direct solution procedures of the partial differential
equations that determine the dynamics of wave functions, the procedure
used here has the advantage of directly yielding expressions for
expectation values and other, in principle observable quantities. The
usual, but often numerically cost-intensive two-step procedure of
first finding expressions for the wave function and then performing
integrations to obtain expectation values is thus significantly
reduced. In this context, the methods used and described are useful
not just for quantum cosmology, the realm of our examples, but may
find wider applicability. In quantum cosmology, or more generally
constrained systems, other advantages include a possible extension to
systems in which the dynamics is determined by a Hamiltonian
constraint, not a Hamiltonian function
\cite{EffCons,EffConsRel,EffConsComp}, and a simplification of the
problem of time in semiclassical regimes
\cite{EffTime,EffTimeLong}.

The semiclassical approximation allows one to consider most cases of
interest in current quantum-gravity research, but they do not allow
one to tackle all of them. For this reason it is of interest to
analyze the system of equations obtained as more and more of the
moments grow large and become relevant when a dynamical state turns
more highly quantum. Since such large numbers of variables coupled to
each other by long evolution equations are difficult to handle
analytically (see the Appendix for deterrence), the main aim of this
article is to provide results of efficient computer-algebra codes
designed to derive these equations automatically and quickly to high
orders, which can then be fed into a numerical solver of the coupled
differential equations. In this way, one can see how much high orders
of the expansions matter and how higher moments can affect the
evolution of expectation values as strong quantum regimes are
approached.

\begin{figure}[ht]
\begin{center}
\includegraphics[width=0.5\textwidth]{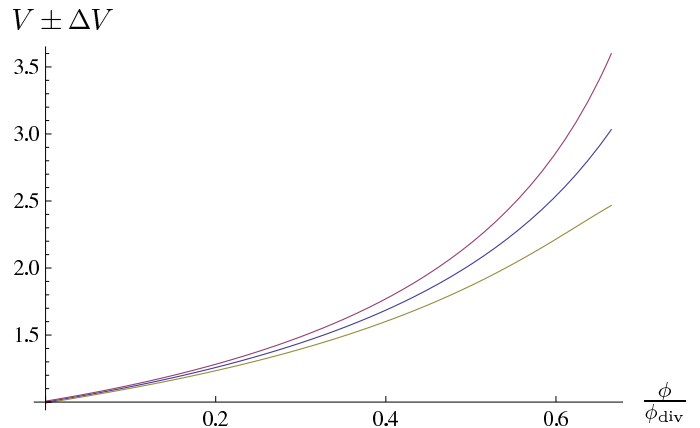}
\caption{\small In this plot we show the evolution of the
volume expectation value $V$ 
for a cosmological constant $\Lambda=9\times 10^7$ and initial data to be 
presented in Sec.~\ref{s:HighOrder}, together with the spread around this 
trajectory $V\pm \Delta V$. This plot illustrates the growing dispersion 
of a state evolving to large volume. The range for $\phi$ shown, relative 
to the value $\phi_{\rm div}$ at which the classical volume diverges, is 
the maximal one obtained with numerical stability to tenth order of 
moments of the evolving state. As the volume grows, the area swept out by 
the wave packet more and more aligns itself with the vertical axis, such 
that volume fluctuations (as well as some other moments) computed at a 
fixed value of $\phi$ diverge. This divergence would happen even if the 
state remained sharply peaked transversal to the direction in which it 
moves, and is thus not a failure of semiclassicality in this system but 
rather an artifact of the parametrization of evolution. Physical 
implications are thus difficult to draw, but an interesting test case for 
back-reaction issues is obtained.
\label{f:VolDiverge}}
\end{center}
\end{figure}

An interesting toy example for this is a spatially flat isotropic
universe with a positive cosmological constant, filled with a
homogeneous free and massless scalar field. The free scalar can play
the role of a global internal time, allowing one to reformulate the
constrained system as ordinary Hamiltonian evolution. In the presence
of a positive cosmological constant, the value of the scalar field is
bounded from above as a function of time; thus, when using the scalar
as internal time, infinite volume is reached at a finite time
$\phi_{\rm div}$. As this point is approached, volume fluctuations and
higher moments diverge even though the regime is supposed to be one of
low-curvature nature (see Figure \ref{f:VolDiverge}). The
non-semiclassical appearance is an artifact of the choice of time,
highlighting the toy-nature of the model. Physical conclusions derived
from the model for a late-time universe with a positive cosmological
constant can thus not be reliable, as interesting as they might
be. But the model is ideal for a mathematical analysis of the possible
roles of large moments. In the present article, we will use this
example mainly to illustrate the usefulness of the numerical codes;
further analysis requires more sophisticated methods to handle the
large parameter space of all moments.

\section{General formalism}

We begin with a brief review of the procedure of effective equations,
followed by a correction of an important formula of \cite{EffAc} and
its accompanying proof.

\subsection{Quantum state space}

Classically, the phase space of a system with one degree of freedom is
completely described by the generalized position $V$ (related to the
volume in cosmological models) and its conjugate momentum $P$ (related
to the Hubble parameter). During the quantization process each
classical phase-space function is promoted to an operator, and
observable information is extracted via expectation values which can
be seen as functionals on the algebra of operators that characterize
states. Since expectation values of products of operators in general
differ from products of expectation values, the following infinitely
many moments are needed to describe the system completely:
\begin{equation}
G^{a,b}:=\langle(\hat P - P)^a \,(\hat V - V)^b\rangle_{\rm Weyl},
\end{equation}
where the subscript Weyl stands for totally symmetric
ordering. Moments cannot be chosen arbitrarily but are subject to
additional conditions such as the uncertainty relation
\begin{equation}\label{uncert}
 G^{2,0}G^{0,2}-(G^{1,1})^2\geq \frac{\hbar^2}{4}\,.
\end{equation} Moreover, a set
of moments satisfying all those conditions in general corresponds to a
mixed rather than pure state. By considering moments to describe
states, mixing (which is important for generic descriptions of
homogeneous quantum cosmology seen as an averaged description of
microscopic physics, in particular inhomogeneities) is automatically included.

The use of moments allows a very general definition of semiclassical
states as those satisfying the $\hbar$ierarchy
$G^{a,b}=O(\hbar^{(a+b)/2})$, realized for Gaussian states (see below)
but also for a much wider class. This characterization of
semiclassical states (pure or mixed) has two major advantages: (i) It
is a general characterization, as unbiased as possible. (In more
restrictive classes such as Gaussians the specific form of the wave
function may actually matter for physical effects.) (ii) The
infinitely many $G^{a,b}$ are decomposed in finite sets to order
$\hbar^n$. This feature allows systematic approximations of
semiclassical dynamics.

The moments are pure quantum degrees of freedom; they would all vanish
in a classical limit. In a quantized interacting system they couple to
the expectation values $P:=\langle\hat{P}\rangle$ and
$V:=\langle\hat{V}\rangle$. The state with respect to which these
moments are taken is specified via values for all the quantum
variables, which are all dynamical and satisfy equations of motion. In
order to consider the evolution of expectation values, the evolution
of all moments must be followed simultaneously. Assuming the moments
to be constant or to take specific values such as, for instance, those
of a Gaussian state at all times cannot provide the correct
dynamics. The dynamics is described by a quantum Hamiltonian, the
expectation value of the Hamiltonian operator, which is state
dependent and thus a function of expectation values and all the
moments. By Taylor expansion, one can write the quantum Hamiltonian as
\begin{equation} \label{HQ}
H_Q(V,P,G^{a,b})=\langle\hat{H}\rangle
= H +\sum_{a=0}^\infty\sum_{b=0}^\infty \frac{1}{a!b!}
\frac{\partial^{a+b} H}{\partial P^a\partial V^b} G^{a,b},
\end{equation}
with the classical Hamiltonian $H$.  The quantum variables it contains
provide crucial corrections to the classical dynamics in a
state-dependent way. We note that expansions of this form (as
also the Feynman or WKB expansions) are expected to be asymptotic, not
necessarily convergent. In the model discussed below, we will
nevertheless observe strong convergence properties within a large
range of orders.

The evolution generated by the quantum Hamiltonian $H_Q$,
governed by the equations of motion ${\rm d}f/{\rm d}\phi=\{f,H_Q\}$ for
phase-space functions $f$, is described with respect to a time
parameter $\phi$. This could be an absolute time parameter or, as in
the example used below, an internal time variable chosen for a
deparameterizable constrained system. (For the effective treatment of
non-deparameterizable constrained systems, lacking a global internal
time, see \cite{EffTime}.)  Effective quantum evolution then provides
an approximation scheme for time-dependent expectation values of
operators, starting with some initial state specified by the initial
expectation values and moments. It is straightforward to find moments
corresponding to a wave-function representation of the initial state
in a Hilbert space, such as the Schr\"odinger one. Moments evolved by
the quantum Hamiltonian then correspond to the dynamical state,
evolved in the Schr\"odinger or Heisenberg picture. We are thus
guaranteed that at all times, as long as the evolved moments remain
finite, there is still a corresponding wave function with those moments to within
the order of approximation considered. It may be difficult to
construct such a wave function explicitly, but this is not required
because the information relevant for observations is already contained
in the expectation values and moments. (When moments diverge, as in the
example discussed below, the state moves out of the domain of
definition of the operators considered. By our moment expansions we
will only consider the approach to such a point of divergence, not the
point itself.)

In order to obtain all Hamiltonian equations of motion in the context
of this effective approach, we will need to compute the Poisson
brackets between any two moments. If one defines the Poisson bracket
for expectation values of arbitrary operators $\hat X$ and $\hat Y$ by
the relation $\{\langle\hat X\rangle,\langle\hat Y\rangle\}=
-i\hbar^{-1}\langle[\hat X,\hat Y]\rangle$ (extended to arbitrary
functions of expectation values using linearity and the Leibniz rule),
Hamiltonian equations of motion generated by
$H_Q=\langle\hat{H}\rangle$ are equivalent to the Schr\"odinger flow
of states or the Heisenberg flow of operators. By applying this
relationship to all powers of the basic operators, the Poisson
brackets between moments is obtained. Since $[\hat V,\hat P]=i\hbar$,
in the particular case of the basic operators themselves, the Poisson
bracket reduces to the classical one, $\{V,P\}=1$.  Moreover, one can
easily show that moments have vanishing Poisson brackets with the
basic expectation values:
\begin{equation}
 \{G^{a,b},V\}=0=\{G^{a,b},P\}\,.
\end{equation}

The general case of Poisson brackets between moments is more involved.
We have the general formula
\begin{eqnarray}\label{GGbrackets}
\{G^{a,b},G^{c,d}\}&=&
a\, d \,G^{a - 1, b} \,G^{c, d - 1} - b \,c\, G^{a, b - 1}\, G^{c - 1, d}
\nonumber\\
&+&\sum_{n}
\left(\frac{i\hbar}{2}\right)^{n-1}
K_{abcd}^{n}\, G^{a+c-n, b+d -n},
\end{eqnarray}
where the sum over $n$ runs over all odd numbers from $1$
to ${\rm Min}(a + c, b + d, a + b, c + d)$, which makes
the coefficients real in spite of the presence
of the imaginary unit. We have also defined
\begin{eqnarray}
K_{abcd}^n &:=&
\sum_{m= 0}^{n}
(-1)^m
m!(n-m)!
\left(\!\!\begin{array}{c}
a\\m
\end{array}\!\!\right)
\left(\!\!\begin{array}{c}
b\\n-m
\end{array}\!\!\right)\nonumber\\&&\!\!\!\!\!\!\!\!\!\!
\times
\left(\!\!\begin{array}{c}
c\\n-m
\end{array}\!\!\right)
\left(\!\!\begin{array}{c}
d\\m
\end{array}\!\!\right)=2^n\sum_{m=0}^n (-1)^m C_{ad}^m C_{bc}^{n-m},
\end{eqnarray}
where the $C$-coefficients come from the Weyl ordering (see also
Sec.~\ref{s:AlgImp}) and are defined as
\begin{equation} \label{C}
C_{kn}^{d}:= \frac{n!\,k!}{(n-d)!(k-d)!(2d)!!}=
\frac{d!}{2^d}
\left(\begin{array}{c}
n\\d
\end{array}\right)
\left(\begin{array}{c}
k\\d
\end{array}\right)\,.
\end{equation}
Note that these coefficients have the permutation property
$K_{abcd}^n=(-1)^n K_{cdab}^n$.  Since the sum in (\ref{GGbrackets})
is only over odd $n$, this makes the antisymmetry of the Poisson
brackets transparent. (In these formulas, $G^{0,0}=1$,
$G^{0,1}=G^{1,0}=0$, and $G^{a,b}=0$ for any $a$ or $b< 0$ are
understood.)

A general formula analogous to (\ref{GGbrackets}) was obtained in
\cite{EffAc}, but since it was not applied to moments of high
orders, a typo remained undiscovered. This typo is corrected in
(\ref{GGbrackets}), and since the formula plays an important role for
the results of this article, we provide a detailed proof in the
following subsection.

\subsection{Poisson structure of moments}

As in \cite{EffAc}, we consider the generating function,
\begin{eqnarray}
&&D(\alpha) = \langle e^{\alpha_P(\hat P - P) + \alpha_V(\hat V - V)}\rangle \nonumber\\
&&\,\,\,\,= \sum_{j=0}^\infty \sum_{k=0}^j \frac{1}{j!} {j \choose k} \langle 
\alpha_P^{j-k} \alpha_V^{k} [(\hat P - P)^{j-k} 
(\hat V - V)^{k}]_{\text{Weyl}} \rangle \nonumber\\
&&\,\,\,\,=  \sum_{j=0}^\infty \sum_{k=0}^j \frac{1}{k!(j-k)!} \alpha_P^{j-k} 
\alpha_V^{k} G^{j-k,k} \,.
\end{eqnarray}
Assuming that the wave function is analytic in the mean values,
$D(\alpha)$ and the Poisson bracket between two of these functions
will also be analytic, which allows us to take the Taylor expansion to
all orders in $\alpha$,
\begin{eqnarray}
\{ D(\alpha), D(\beta)\} &=& \sum_{w=0}^\infty \sum_{x=0}^w \sum_{y=0}^\infty 
\sum_{z=0}^y \frac{\alpha_P^{w-x}\alpha_V^x \beta_P^{y-z}
\beta_V^z}{x!(w-x)!z!(y-z)!}\nonumber\\
&&\times\{G^{w-x,w},G^{y-z,y}\}\,.
\end{eqnarray}

On the other hand, we use the Baker-Campbell-Hausdorff formula
to get the commutator
\begin{eqnarray}
[ e^{\alpha_P\hat P + \alpha_V\hat V},  e^{\beta_P\hat P+ \beta_V\hat V}] &=& 2i \sin \left( \frac{\hbar}{2}(\alpha_V \beta_P - \alpha_P \beta_V)\right)\nonumber\\&&\times\,\, e^{(\alpha_P+\beta_P)\hat P + (\alpha_V+\beta_V)\hat V},
\end{eqnarray}
which gives us the Poisson bracket
\begin{eqnarray}
\{ D(\alpha), D(\beta)\} &=& \frac{2}{\hbar}\sin \left( \frac{\hbar}{2} (\alpha_V \beta_P - \alpha_P \beta_V)\right) D(\alpha + \beta)
\nonumber\\ \label{DaDb}
&-& (\alpha_V \beta_P - \alpha_P \beta_V)D(\alpha)D(\beta).
\end{eqnarray}
We then use a Taylor series and the binomial theorem to expand the right-hand side
of the last equation. For the first term we get the expression,
\begin{eqnarray}
&&\frac{2}{\hbar}\sin \left( \frac{\hbar}{2} (\alpha_V \beta_P - \alpha_P \beta_V)\right) D(\alpha + \beta)
\nonumber\\
&=& \frac{2}{\hbar} \bigg[ \sum_{r=0}^\infty \left(\frac{\hbar}{2}\right)^{2r+1} \frac{(-1)^r}{(2r+1)!}
\nonumber\\
&&\times\sum_{s=0}^{2r+1}  {2r+1 \choose s}(\alpha_V\beta_P)^{2r+1-s} (-\alpha_P\beta_V)^s  \bigg] D(\alpha+\beta)
\nonumber\\
&=& \sum_{r=0}^{\infty}\sum_{s=0}^{2r+1}\sum_{j=0}^\infty\sum_{k=0}^j \sum_{m=0}^{j-k} \sum_{n=0}^{k} \left(\frac{\hbar^2}{4}\right)^r \frac{(-1)^{r+s}}{s!(2r+1-s)!}
\nonumber\\
&&   \!\!\!\!\times \frac{\alpha_P^{s+j-k-m}\alpha_V^{2r+1-s+k-n}\beta_P^{2r+1-s+m}\beta_V^{s+n}}{m!(j-k-m)!n!(k-n)!} G^{j-k,k},
\end{eqnarray}
whereas, for the second one, the expansion is given by
\begin{eqnarray}
&&(\alpha_P \beta_V - \alpha_V \beta_P)D(\alpha)D(\beta)
\nonumber\\
&=& \sum_{j=0}^\infty \sum_{k=0}^j \sum_{m=0}^\infty \sum_{n=0}^m \frac{G^{j-k,k}G^{m-n,n}}{k!(j-k)!n!(m-n)!}
\nonumber\\
&&\!\!\!\! \!\!\!\!\!\!\!\!\!\!\times\Big(\alpha_P^{j-k+1}\alpha_V^{k}\beta_P^{m-n}\beta_V^{n+1} - \alpha_P^{j-k}\alpha_V^{k+1}\beta_P^{m-n+1}\beta_V^{n}\Big).
\end{eqnarray}
Since (\ref{DaDb}) must be satisfied for all values of $\alpha$
and $\beta$ we find that the coefficients in front of each of the
powers of various $\alpha$'s and $\beta$'s must be equal. By direct
comparison we then obtain the Poisson bracket between our quantum variables.
Making the convenient substitutions $a=w-x$, $b=x$, $c=y-z$, and $d=z$,
and combining the factorials into binomial coefficients the result reads,
\begin{eqnarray}
\{G^{a,b},G^{c,d}\} &=&
adG^{a-1,b}G^{c,d-1} - bcG^{a,b-1}G^{c-1,d}
\nonumber\\&+&
\sum_{r=0}^\infty \sum_{s=0}^{2r+1}
2 \hbar^{2 r} (-1)^{r+s} C_{ad}^s C_{bc}^{2r+1-s}
\nonumber\\&&\qquad\times\,
G^{a+c-2r-1,b+d-2r-1}.\,\,
\end{eqnarray}
Finally we make the replacement $n=2r+1$, noting that in the
summation over $r$ the only non-zero terms occur when $2r+1 < {\rm
Min}(a + c, b + d, a + b, c + d)$,
which gives us (\ref{GGbrackets}).

\section{Algebraic implementation}
\label{s:AlgImp}

\begin{figure}[h]
\begin{center}
\includegraphics[width=0.5\textwidth]{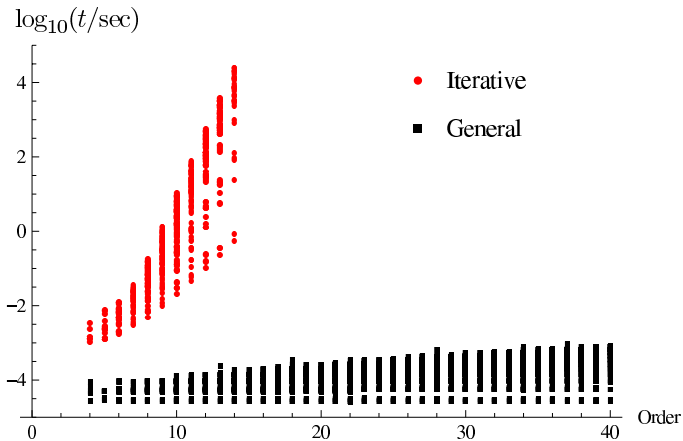}
\caption{\small In this figure we compare the timing of computing the
Poisson brackets between any two moments with the code working
iteratively as well as the one that makes use of the general formula
(\ref{GGbrackets}). The time, shown in a logarithmic scale (to base
10), has been measured in seconds. The order plotted in the graphics
corresponding to a bracket $\{G^{a,b},G^{c,d}\}$ has been defined as
the sum of all the indices: $a+b+c+d$.  Note the exponential increase
of the timing for the recursive code, whereas the computations with
the general code are kept in the order of $10^{-4}$ seconds up to the
considered order.  There are $N\equiv(n+1)(n+2)/2-3$ different moments
with an order less or equal to $n$, where the subtraction of 3 comes
from the fact that the order 0 and 1 moments are trivial; they can be
combined into $N(N+1)/2$ different Poisson brackets.  In the case of
the recursive (general) code, the 33 (228) moments up to order seven
(20) have been considered, and in the plot the timing for their
corresponding 561 (26106) independent Poisson brackets are drawn.}
\label{timing}
\end{center}
\end{figure}

We have written two different codes in \emph{Mathematica} in order to obtain
the Poisson brackets between any two generic moments. The first code
works iteratively, whereas the second one makes use of the general
formula (\ref{GGbrackets}). The results agree for all orders checked
(up to 14th) but the second code is much more efficient. This can be
clearly seen in Fig. \ref{timing}, where we have plotted the timing
of both codes to compute several Poisson brackets. The
timing of the recursive code increases exponentially with the order
of the considered Poisson bracket, whereas the general code allows
us to obtain high-order results in a very short time. This fact
shows the importance of the general formula (\ref{GGbrackets})
from a practical point of view.

In the following we briefly explain how the recursive code works.
Making use of the binomial identity,
the expression for the moments can be written as,
\begin{eqnarray}
G^{a,b}&=&\sum_{k=0}^{a}\sum_{n=0}^{b}
(-1)^{a+b-k-n}
\left(\begin{array}{c}
a\\k
\end{array}\right)
\left(\begin{array}{c}
b\\n
\end{array}\right)
\nonumber\\
&&\times P^{a-k}V^{b-n}
\langle\hat P^k\,\hat V^n\rangle_{\rm Weyl}.
\end{eqnarray}
On the other hand,
we also have found that the Weyl ordering can be
given as a linear combination of expectation values
with a predefined order, for any operators $\hat A$
and $\hat B$,
\begin{equation}
\langle\hat A^k\,\hat B^n \rangle_{\rm Weyl}\,=
\sum_{d=0}^{{\rm Min}(k,n)}[\hat B, \hat A]^d\, C_{kn}^{d}\, \langle\hat 
A^{k-d} \hat B^{n-d}\rangle,
\end{equation}
where the coefficients $C_{kn}^{d}$ have been defined in (\ref{C}).
Hence, for our case of interest, we can write either
\begin{equation}
\langle\hat P^k\,\hat V^n \rangle_{\rm Weyl}\,=
\sum_{d=0}^{{\rm Min}(k,n)}(i\hbar)^d C_{kn}^{d} \langle\hat P^{k-d} 
\hat V^{n-d}\rangle,
\end{equation}
or,
\begin{equation}
\langle\hat P^k\,\hat V^n \rangle_{\rm Weyl}\,=
\sum_{d=0}^{{\rm Min}(k,n)}(-i\hbar)^d C_{kn}^{d} \langle\hat V^{n-d}
\hat P^{k-d}\rangle.
\end{equation}
In order to obtain the Poisson brackets between
different moments iteratively we make use of
the definition of the moments in the form,
\begin{eqnarray}
G^{a,b}&=&\sum_{k=0}^{a}\sum_{n=0}^{b}\sum_{d=0}^{{\rm Min}(k,n)}
(-i\hbar)^d (-1)^{a+b-k-n}
\left(\begin{array}{c}
a\\k
\end{array}\right)
\left(\begin{array}{c}
b\\n
\end{array}\right)
\nonumber\\\label{defG}
&&\times C_{kn}^{d}
P^{a-k}V^{b-n}
\langle \hat V^{n-d}\hat P^{k-d}\rangle.
\end{eqnarray}
For computing $\{G^{a,b},G^{c,d}\}$ we then use the properties of the Poisson
brackets as well as of the commutator until all operator terms reduce
to the simple commutator between the isolated operators $\hat V$ and
$\hat P$.

\section{A massless scalar field with cosmological constant}

As an application of high-order quantum back-reaction we now introduce
a cosmological model with a regime of large moments: an isotropic
spatially flat universe filled with a free, massless scalar $\phi$
(with momentum $p_{\phi}$) and a positive cosmological constant
$\Lambda>0$.  From the Friedmann equation
\begin{equation}
 \left(\frac{a^\prime}{a}\right)^2= \frac{4\pi
 G}{3}\frac{p_{\phi}^2}{a^6}+\Lambda,
\end{equation}
we obtain, ignoring constant factors for the sake of compactness,
$p_{\phi}=a^2 \sqrt{a^{\prime 2}-\Lambda a^2}$ where the prime is the
derivative with respect to cosmological time.  The square root
may take both signs, corresponding to expanding and contracting
cosmological solutions. We will choose the positive sign in order to
analyze expanding universe models approaching large volume, as they remain
in a low-curvature regime. The opposite sign, if desired, could be
easily included since it would just invert the sign of the constant of
motion $p_{\phi}$ [which will be the deparameterized classical Hamiltonian
of our system (\ref{classH})] as
well as the effective Hamiltonian (\ref{ourhamiltonian}) below.
Therefore, from
the expanding solutions we show it will be straightforward to get their
contracting counterparts just by performing a time reversal.

We rewrite $p_{\phi}$ as
\begin{equation}\label{pphi}
p_{\phi}= (1-x)V\sqrt{P^2-\Lambda[(1-x)V]^{(1+2x)/(1-x)}}
\end{equation}
in terms of canonical gravitational variables $V=(1-x)^{-1}a^{2-2x}$
and $P=-a^{2x}a^{\prime}$. The parameter $x$ characterizes different
cases of lattice refinement of an underlying discrete state; see
\cite{InhomLattice,CosConst}. The quantum dynamics (Wheeler-DeWitt
type) that will be implemented in this paper is insensitive to the
change of this discreteness parameter. However, some values lead
to advantages in the solution procedure, as we will make use of
below. Moreover, it is useful to keep the parameter in the definition
of canonical variables in order to facilitate possible applications
to loop quantum cosmology in the near future, where this parameter
will become relevant.

The system is simplest to deal with for
$x=-1/2$, in which case $V$ is proportional to the volume, because
$p_{\phi}$, which can be thought of as the Hamiltonian for evolution
in internal time $\phi$, is then linear in $V$. For a negative
cosmological constant, the effective evolution coming from a
quantization of (\ref{pphi}) has been solved and analyzed in
\cite{Recollapse}, some of whose formulas apply here as well. (For
numerical wave-function evolution of this model with $\Lambda<0$, see
\cite{NegCosNum}.) For $\Lambda<0$, however, the moments do not grow
large at large volume, and the volume is bounded by recollapse. These
two features are not present for $\Lambda>0$, making an analysis of
high-order moments necessary, the topic of this article.

From now on we use $x=-1/2$, which implies
\begin{equation}\label{classH}
 p_{\phi}=H= \frac{3}{2}V\sqrt{P^2-\Lambda}
\end{equation}
as the Hamiltonian of the deparameterized system.
Since the dependence of the classical Hamiltonian on $V$ is linear, we
will only obtain terms of the form $G^{a,0}$ and $G^{a,1}$ in the
quantum Hamiltonian and we have the closed expression,
\begin{eqnarray}
H_Q&=& \frac{3}{2}V\sqrt{P^2-\Lambda}+ \frac{3}{2}\sqrt{\Lambda}
\sum_{n=2}^{\infty} \frac{\Lambda^{-n/2}}{n!}\bigg[ V T_{n}(P/\sqrt{\Lambda}) G^{n,0}
\nonumber\\\label{ourhamiltonian}
&&\qquad\qquad+ \,n\,\sqrt{\Lambda}\, T_{n-1}(P/\sqrt{\Lambda}) G^{n-1,1}\bigg]\,,
\end{eqnarray}
where we have defined
\begin{equation}
 T_n(x):= \frac{\md^n}{\md x^n} \sqrt{x^2-1}.
\end{equation}
Explicitly, the moments appearing here are
\begin{eqnarray*}
 G^{n,0} &:=& \langle(\hat{P}-\langle\hat{P}\rangle)^n\rangle,\\
 G^{n-1,1} &:=& \frac{1}{n}\sum_{i=0}^{n-1}\langle
 (\hat{P}-\langle\hat{P}\rangle)^i (\hat{V}-\langle\hat{V}\rangle)
 (\hat{P}-\langle\hat{P}\rangle)^{n-1-i}\rangle\,.
\end{eqnarray*}
Hamiltonian (\ref{ourhamiltonian}) will describe the semiclassical
behavior of the Wheeler-DeWitt quantization of this system.

Equations of motion for $V$ and $P$ are derived as in any Hamiltonian
system, noting that the expectation values Poisson commute with any
quantum variable. This gives
\begin{eqnarray}
\dot{V} &=& \frac{3}{2}V\frac{P}{\sqrt{P^2-\Lambda}} + \frac{3}{2}
\sum_{n=2}^{\infty} \frac{\Lambda^{-n/2}}{n!}\bigg[ V T_{n+1}(P/\sqrt{\Lambda}) G^{n,0}
\nonumber\\\label{dotV}
&&\qquad\qquad +\, n \,\sqrt{\Lambda}\, T_{n}(P/\sqrt{\Lambda}) G^{n-1,1}\bigg]\,,\\\label{dotP}
\dot{P}
&=& -\frac{3}{2}\sqrt{P^2-\Lambda}- \frac{3}{2}\sqrt{\Lambda}
\sum_{n=2}^{\infty} \frac{\Lambda^{-n/2}}{n!}
T_n(P/\sqrt{\Lambda}) G^{n,0},\qquad
\end{eqnarray}
where dot stands for derivatives with respect to $\phi$
and we have made use of the fact that
\begin{equation}
\left\{V,T_n(P/\sqrt{\Lambda})\right\}=\frac{1}{\sqrt{\Lambda}}T_{n+1}(P/\sqrt{\Lambda}).
\end{equation}

Equations of motion for the quantum variables can also be derived from
Poisson relations between them
\begin{eqnarray}
\dot{G}^{r ,s } &=&\frac{3}{2}\sqrt{\Lambda}
\sum_{n=2}^{\infty} \frac{\Lambda^{-n/2}}{n!}\bigg[ V T_{n}(P/\sqrt{\Lambda}) \left\{ G^{r,s},G^{n,0} \right\}
\nonumber\\\label{generalG}
&&\qquad +n\sqrt{\Lambda} T_{n-1}(P/\sqrt{\Lambda}) \left\{ G^{r,s},G^{n-1,1} \right\} \bigg].
\end{eqnarray}
In fact, this highly coupled system of infinitely many variables
is extremely complicated to solve, and we will try to analyze it by
means of truncations to finitely many variables which include the
expectation values and only low order quantum variables.

\subsection{Vanishing $\Lambda$}\label{L0}

For $\Lambda=0$, we obtain the harmonic model of \cite{BouncePert} for
which all equations decouple:
\begin{eqnarray}
\dot V &=& \frac{3}{2} V,\\
\dot P &=& -\frac{3}{2} P,\\
\dot{G}^{ab} &=& \frac{3}{2} (b-a) G^{ab}\,.
\end{eqnarray}
There is no quantum back-reaction, and
all equations can easily be solved:
\begin{eqnarray}
V(\phi)&=&V_0 \exp{\left[\frac{3}{2} (\phi-\phi_0)\right]},\\
P(\phi)&=&P_0 \exp{\left[-\frac{3}{2} (\phi-\phi_0)\right]},\\
G^{ab}(\phi)&=&G^{ab}_0 \exp{\left[\frac{3}{2} (b-a) (\phi-\phi_0)\right]}
\label{GabSol}\,.
\end{eqnarray}

\subsection{Classical equations}

To classical order, i.e., assuming all moments to vanish, we obtain
analytical solutions
\begin{widetext}
\begin{eqnarray}
P_{\rm classical}(\phi) &=& P_0 \cosh(3(\phi-\phi_0)/2)-
\sqrt{P_0^2-\Lambda} \sinh(3(\phi-\phi_0)/2)\,,\label{Pclass}\\
V_{\rm classical}(\phi) &=& V_0\frac{\sqrt{P_0^2-\Lambda}}{\sqrt{P_0^2-\Lambda}
\cosh(3(\phi-\phi_0)/2)-P_0\sinh(3(\phi-\phi_0)/2)}\,,\label{Vclass}
\end{eqnarray}
\end{widetext}
where $V_0$ and $P_0$ are the value of the expectation values at the
initial time $\phi_0$.
At the time when $\phi_{\rm div}:=\phi_0+2/3 \tanh^{-1}
(\sqrt{P_0^2-\Lambda}/P_0)$ the volume $V_{\rm classical}(\phi)$
diverges, and $\sqrt{P_{\rm classical}(\phi_{\rm div})^2-\Lambda}=0$.

\subsection{Second order}\label{2ndorder}

To understand whether quantum effects can be important in this
large-volume regime of small curvature (if $\Lambda$ is small), we
will analytically solve (under certain approximations) the equations
of motion for second order quantum variables,
i.e.\ fluctuations $G^{2,0}$, $G^{0,2}$ and the covariance $G^{1,1}$. To
this order, ignoring higher moments, we have a quantum Hamiltonian
\begin{eqnarray}
 H_Q &=& \frac{3}{2}V\sqrt{P^2-\Lambda}- \frac{3}{4}\Lambda
\frac{V}{(P^2-\Lambda)^{3/2}} G^{2,0}\nonumber\\&+&
\frac{3}{2}\frac{P}{\sqrt{P^2-\Lambda}} G^{1,1} \,.
\end{eqnarray}
Using the Poisson bracket relations
\begin{eqnarray}
&&\{G^{0,2},G^{1,1}\}= 2G^{0,2},\quad\quad \{G^{0,2},G^{2,0}\}=
 4G^{1,1},
\nonumber\\ \label{PoissonSecond} && \{G^{1,1},G^{2,0}\}=2G^{2,0}\,,
\end{eqnarray}
the moments satisfy the equations of motion,
\begin{eqnarray}
 \dot{G}^{2,0} &=& -3\frac{P}{\sqrt{P^2-\Lambda}} G^{2,0} \,,
\label{GPPeom}\\
 \dot{G}^{1,1} &=& -\frac{3}{2}\Lambda\frac{V}{(P^2-\Lambda)^{3/2}}
 G^{2,0}\,,
\label{GVPeom}\\
 \dot{G}^{0,2} &=& -3 \Lambda\frac{V}{(P^2-\Lambda)^{3/2}} G^{1,1}
 +3\frac{P}{\sqrt{P^2-\Lambda}} G^{0,2}\,. \label{GVVeom}
\end{eqnarray}

\subsubsection{Expectation values as background solutions}

As a first approximation, we can use the classical solutions for $V$
and $P$ as some kind of background on which the quantum variables
evolve. From the resulting solutions we then determine whether
quantum back-reaction effects by quantum variables on expectation
values are significant.

The equations for fluctuations to this level of truncation are not
strongly coupled, which allows us to solve for all of them by
integrations. For $G^{2,0}$, we obtain
\begin{eqnarray}
G^{2,0}(\phi) &=& G^{2,0}(\phi_0)\bigg{(}\cosh(3(\phi-\phi_0)/2)
\nonumber\\\label{GPPsol}
&-&\frac{P_0}{\sqrt{P_0^2-\Lambda}} \sinh(3(\phi-\phi_0)/2)\bigg{)}^2\,,
\end{eqnarray}
which then allows us to solve for $G^{1,1}$ and, in turn,
$G^{0,2}$. We do not present these solutions here (which can be found
from \cite{Recollapse} by changing the sign of $\Lambda$), but note
that (\ref{GVVeom}) can be written as
\begin{equation}
 \frac{\md}{\md\phi} \left(\frac{G^{0,2}}{V_{\rm classical}^2}\right) = -3\Lambda
 \frac{G^{1,1}}{V_{\rm classical}(P^2-\Lambda)^{3/2}}\,.
\end{equation}
(Within the second-order approximation, we could replace $V_{\rm
classical}$ with $V$ in this equation up to terms of the form
$G^{0,2}G^{2,0}$ and $G^{0,2}G^{1,1}$.)
This shows that the evolution of volume
fluctuations relative to volume is determined by the covariance
$G^{1,1}$. For an unsqueezed state, we would have $G^{1,1}=0$ and
squared volume fluctuations would be proportional to the total
volume. However, even if we start with an unsqueezed initial state,
squeezing would develop over time since $\dot{G}^{1,1}$ is
non-zero. In this way, the precise state has to be analyzed to
understand the long-term evolution of fluctuations.

\subsubsection{Coupled equations}

The solution (\ref{GPPsol}) for $G^{2,0}$, compared with (\ref{Vclass}),
shows that curvature fluctuations to second order vanish where $V$
diverges. Thus, also $G^{0,2}$ has to diverge there in order to
respect the uncertainty relation. The approach to zero is the same for
$\sqrt{G^{2,0}}$, $1/V$ and $\sqrt{P^2-\Lambda}$, which we can use to
check whether quantum back-reaction is important. Writing equations of
motion for expectation values to the order where coupling terms to
fluctuations appear, we obtain
\begin{eqnarray}
 \dot{P} &=& -\frac{3}{2}\sqrt{P^2-\Lambda}+ \frac{3}{4}\Lambda
 \frac{G^{2,0}}{(P^2-\Lambda)^{3/2}}\,, \label{Peom}\\
 \dot{V} &=& \frac{3}{2} \frac{VP}{\sqrt{P^2-\Lambda}}+
 \frac{9}{4}\Lambda \frac{VP G^{2,0}}{(P^2-\Lambda)^{5/2}}
\nonumber\\
&-&\frac{3}{2}\Lambda \frac{G^{1,1}}{(P^2-\Lambda)^{3/2}}\,, \label{Veom}
\end{eqnarray}
which now forms a coupled set of five differential equations if we
include $G^{0,2}$. When $V$ diverges, coupling terms due to quantum
back-reaction cannot be ignored. In $\dot{P}$, for instance, the
coupling term diverges, and in fact dominates the classical term,
since $G^{2,0}$ vanishes like $P^2-\Lambda$, and also $\dot{V}$
has diverging coupling terms. Even the truncation used here, which
does include some quantum corrections, is not consistent at this point
because one has to expect that further correction terms by higher
moments become large, too. Nevertheless, in the approach to the
divergence of $V$, correction terms should become relevant one by one,
depending on their moment order. Thus, it is of interest to analyze
the coupled system written here to second order in the moments before
proceeding to higher orders.

We first focus on the system given by the equations (\ref{Peom}) for
$\dot{P}$ and (\ref{GPPeom}) for $\dot{G}^{2,0}$ since these two
equations are coupled only with each other. If we divide $\dot{P}$ by
$\dot{G}^{2,0}$, we obtain the differential equation
\begin{equation}
\frac{\md P}{\md G^{2,0}} = \frac{1}{2}\frac{P^2-\Lambda}{PG^{2,0}}-
\frac{1}{4}\frac{\Lambda}{P(P^2-\Lambda)},
\end{equation}
for $P(G^{2,0})$ or, in a simpler form,
\begin{equation}
 \frac{\md (P^2-\Lambda)}{\md \log G^{2,0}} = P^2-\Lambda-
 \frac{1}{2}\frac{\Lambda}{P^2-\Lambda} G^{2,0}\,.
\end{equation}
This equation of the form
\[
 f'(x) = f(x)- \frac{1}{2}\Lambda\frac{e^x}{f(x)}
\]
can be solved by
\[
 f(x)=\sqrt{ce^{2x}+\Lambda e^x}
\]
with an integration constant $c$. In this way, we derive the
exact second-order relations
\begin{eqnarray}\label{relationPG1}
 P &=& \sqrt{\Lambda+\sqrt{c (G^{2,0})^2+
     \Lambda G^{2,0}}}\,,\\\label{relationPG}
 \sqrt{P^2-\Lambda} &=& \sqrt[4]{c(G^{2,0})^2+
   \Lambda G^{2,0}}\,.
\end{eqnarray}

We can use these relations to solve the equations of motion for
$P(\phi)$ and $G^{2,0}(\phi)$, but the resulting differential equations
are complicated. For $G^{2,0}$, for instance, the differential equation
is
\begin{equation} \label{GPPphieom}
 \dot{G}^{2,0} = -3\frac{P}{\sqrt{P^2-\Lambda}} G^{2,0}= -3\sqrt{(G^{2,0})^2+
   \frac{\Lambda G^{2,0}}{\sqrt{c+\Lambda/G^{2,0}}}}\,.
\end{equation}
Fortunately, as we will see in the next subsection,
already the relation $P(G^{2,0})$ has interesting
implications regarding the quantum back-reaction problem. We will
continue using (\ref{GPPphieom}) in the next subsection, but first
note that further analysis, at least numerically, of the truncated
second order system could be facilitated by an explicit decoupling of
the equations. We have already provided direct equations for $P$ and
$G^{2,0}$ which we can assume to be solved by integrations. Further
equations are then still coupled amongst each other, but different
combinations are decoupled. First, we combine $\dot{V}$ and
$\dot{G}^{1,1}$ to
\begin{eqnarray}
 \frac{\md}{\md\phi}  \left(\frac{G^{1,1}}{V}\right) &=& -\frac{3}{2}
 \Lambda \frac{G^{2,0}}{(P^2-\Lambda)^{3/2}}- \frac{3}{2}
 \frac{P}{\sqrt{P^2-\Lambda}} \frac{G^{1,1}}{V}
\nonumber\\ &-& \frac{9}{4}\Lambda
 \frac{P}{(P^2-\Lambda)^{5/2}} G^{2,0}\frac{G^{1,1}}{V}
\nonumber\\ &+&
 \frac{3}{2}\Lambda \frac{1}{(P^2-\Lambda)^{3/2}}
 \left(\frac{G^{1,1}}{V}\right)^2
\end{eqnarray}
which is already decoupled from the rest. Once $G^{1,1}/V$ has been
solved for, (\ref{Veom}) presents an uncoupled differential equation
for $V$, and we can finally integrate (\ref{GVVeom}) for $G^{0,2}$.

\subsubsection{Strength of quantum back-reaction}

The term $\Lambda G^{2,0}$ in (\ref{relationPG}) arises through the
quantum back-reaction in the evolution equation for $P$ (\ref{Peom}).
If this term is ignored, we reproduce the previous approximated
solution $G^{2,0}=G^{2,0}(\phi_0) (P^2-\Lambda)
/(P_0^2-\Lambda)$, which follows from (\ref{Pclass}) and
(\ref{GPPsol}). Hence, comparison shows that
the integration constant is proportional to $c\sim
(P_0^2-\Lambda)^2/[G^{2,0}(\phi_0)]^2$ if initial values are imposed where
back-reaction is weak. Starting with a semiclassical initial state,
for some time we will have $G^{2,0}(\phi)\sim G^{2,0}(\phi_0)$ and the term
$\Lambda G^{2,0}$ can indeed be ignored compared to the classical
parameter $(P_0^2-\Lambda)^2$ in (\ref{relationPG}).
(The value of $P_0^2-\Lambda$ cannot vanish under the stated
assumptions, for we already know that back-reaction terms would then
be large at $\phi_0$.) But as we approach the divergence of $V$, we
have seen that $G^{2,0}$ according to the lowest order solution (\ref{GPPsol}) would
tend to zero. In this regime, the new term $\Lambda/G^{2,0}$ in (\ref{GPPphieom})
will eventually dominate over the constant $c$. This is another
demonstration of the fact that quantum back-reaction is important at
this place.

For $c=0$, i.e. in the strong quantum back-reaction regime where
$\Lambda/G^{2,0}$ dominates, we can find a solution for
(\ref{GPPphieom}). The combination $h(\phi):=\sqrt[4]{G^{2,0}(\phi)/\Lambda}$
satisfies a simple differential equation solved by
$G^{2,0}(\phi)=\Lambda \sinh^4(3(\phi-\phi_0)/4)$. The
higher power compared to the solution (\ref{GPPsol}) in the absence of
back-reaction again indicates that quantum back-reaction becomes
important at large volume.

Turning now to $V$, we can confirm that the quantum back-reaction term
does change its behavior of divergence. In (\ref{Veom}), the dominant
back-reaction term involves $G^{2,0}$ rather than $G^{1,1}$. Ignoring
quantum back-reaction in the solution for $G^{2,0}$, we would have
\[
 \frac{\dot{V}}{V} \sim \frac{3}{2} \frac{P}{\sqrt{P^2-\Lambda}}
 +\frac{9}{4} \frac{\Lambda P}{\sqrt{c}(P^2-\Lambda)^{3/2}}\,.
\]
The second term is subdominant.  Quantum back-reaction in the solution
for $G^{2,0}$, on the other hand, implies a stronger divergence as can
be seen by using relation (\ref{relationPG1}) for $c=0$, since we now have
\[
 \frac{\dot{V}}{V} \sim \frac{15}{4} \frac{P}{\sqrt{P^2-\Lambda}}\,.
\]
Here, the quantum back-reaction term for $V$ has the same form as the
classical term, which have thus be combined.  The pre-factor of the
right-hand side becomes larger than classically once quantum back-reaction
is included, strengthening the divergence of $V$.

One may conjecture that, with all quantum corrections from the
complete series of moments, the divergence of $V$ would
disappear. This is suggested by the fact that a self-adjoint
Hamiltonian would provide unitary evolution in which the wave function
remains well-defined throughout the point where the classical $V$
diverges, and even beyond that point. In the model considered, the
Hamiltonian for $\phi$-evolution is not essentially self-adjoint but
has several inequivalent self-adjoint extensions
\cite{SelfAdFlat}. While subtle issues of self-adjoint extensions of
operators usually do not play a large role in effective equations,
provided that consistent truncations are possible, the point where $V$
diverges classically requires all moments to be taken into account.
With infinitely many independent variables, coupling terms have to be
arranged in a precise manner, which can reflect issues of self-adjoint
extensions. When all moments are taken into account in a way
corresponding to unitary evolution of the corresponding wave function,
well-defined equations for expectation values and the moments may
result. This supports the conjecture that the classical divergence
might be removed by an interplay of all moments, but a verification in
this highly coupled system is complicated. (It could, in fact, happen
that wave functions evolve out of the domain of definition of the
unbounded operator $\hat{V}$, and then directly re-enter. Thus, it is
not obvious that equations for expectation values
$\langle\hat{V}\rangle$ must be well-defined at all times. Another
example which indicates caution against applying arguments of
self-adjointness for the behavior of expectation values is the
upside-down harmonic oscillator. Its Hamiltonian is not essentially
self-adjoint and allows different self-adjoint extensions, depending
on the boundary conditions at infinity where wave packets are
reflected. However, the Hamiltonian is also quadratic, and no quantum
back-reaction results. Effective equations are automatically truncated
without implementing any approximation; the evolution of moments to
finite order just does not depend on what self-adjoint extension is
used.)

We will probe the divergence of $V$ further by the numerics of
high-order moments. For now, conclusions one can draw already from the
analysis presented so far are:
\begin{itemize}
 \item Quantum back-reaction is essential at large volume in the
 presence of a positive cosmological constant, even though one would
 expect it to be a semiclassical regime of small curvature if
 $\Lambda$ is small.
\item While a wave packet in a $V$-representation may not show large
 deviations from the classical trajectory, this is only because volume
 fluctuations grow and are larger than deviations from the classical
 trajectory. Nevertheless, as volume fluctuations diverge, deviations
 from the classical trajectory can be large. Equations including
 quantum back-reaction clearly show that correction terms cannot be
 ignored. Quantum back-reaction seems to strengthen the divergence of
 $V$ rather than triggering a recollapse.
\item Corrections are more visible for curvature $P$ since its
  fluctuations approach zero while also here quantum back-reaction
  terms in the equations of motion are relevant. Thus, deviations from
  the classical trajectory build up and, unlike for $V$, are not
  covered by a broad wave function if a $P$-representation is used.
\item All state parameters matter for the spreading behavior and for
the exact size of quantum corrections. A state may be assumed to be
unsqueezed initially, setting $G^{1,1}(\phi_0)=0$, but correlations will
build up over time.
\end{itemize}

\subsection{Third order}

At third or higher orders, the equations of motion not only
become longer; there is also a new feature in the Poisson relations
which we illustrate here. Some examples of third-order moments are
\begin{eqnarray*}
G^{3,0} &=& \langle\hat{P}^3\rangle-3P\ G^{2,0}-P^3, \\
 G^{2,1} &=& \frac{1}{3} \langle \hat{V} \hat{P}^2+ \hat{P}\hat{V}\hat{P}+ \hat{P}^2 \hat{V}\rangle -2P\ G^{1,1}
\\
&-& V\ G^{2,0}- P^2 \,V, \\
G^{1,2} &=& \frac{1}{3} \langle \hat{P} \hat{V}^2+ \hat{V}\hat{P}\hat{V}+ \hat{V}^2 \hat{P}\rangle -2V\ G^{1,1}
\nonumber\\
&-& P\ G^{0,2}- V^2 P\,.
\end{eqnarray*}
The Poisson brackets of second-order with third-order moments are of
third-order form, for instance:
\begin{eqnarray*}
\nonumber \left\{ G^{0,2},G^{3,0} \right\} = 6 G^{2,1}\quad&,&\quad
\nonumber \left\{ G^{0,2},G^{2,1} \right\} = 4 G^{1,2}\,,\\
\nonumber \left\{ G^{1,1},G^{3,0} \right\} = 3 G^{3,0}\quad&,&\quad
\nonumber \left\{ G^{1,1},G^{2,1} \right\} =  G^{2,1}\,, \\
\left\{ G^{2,0},G^{2,1} \right\} = -2 G^{3,0}\quad&,&\quad
\left\{ G^{2,0},G^{3,0} \right\} = 0,
\end{eqnarray*}
where only the ones needed for evolution in our cosmological system
were written. Two third-order moments in a Poisson bracket, on the
other hand, show a different behavior: They produce a term of higher
order (fourth) as well as products of second-order moments, which are
to be considered as being of fourth order in the $\hbar$ierarchy. This
behavior of the Poisson structure has consequences for the truncation
of equations of motion.

Unlike for second order, (\ref{PoissonSecond}), there is a difference
between expanding the Hamiltonian to third order and then deriving
equations of motion, and expanding the equations of motion (obtained
from an expanded Hamiltonian or the whole series in $\hbar$) to third
order. In particular, moments of third order in the Hamiltonian
produce moments of fourth order in the equations of motion for third
order moments.  This would result in a complete set if also equations
of motion for fourth order moments are included. This problem does not
arise if one first computes equations of motion and expands those to
third (or any) order which is anyway more suitable because it is the
equations of motion which we are primarily interested in. We
illustrate this feature by the following evolution equations, computed
from the third-order Hamiltonian:
\begin{widetext}
\begin{eqnarray}
 \dot{P} &=& -\frac{3}{2} \sqrt{P^2 - \Lambda} + 
  \frac{3}{4} \Lambda \frac{G^{2,0}}{(P^2 - \Lambda)^{3/2}} - 
  \frac{3}{4} \frac{\Lambda  P}{(P^2 - \Lambda)^{5/2}} G^{3,0}\,,\\
\dot{V} &=& 
 \frac{3}{2} \frac{VP}{\sqrt{P^2 - \Lambda}} + 
  \frac{9}{4} \Lambda^2\frac{VPG^{2,0}}{(P^2 - \Lambda)^{5/2}} - 
  \frac{3}{2} \Lambda^2 \frac{G^{1,1}}{(P^2 - \Lambda)^{3/2}} - 
  \frac{3}{4} \Lambda^2\frac{V(4P^2 + \Lambda)}{(P^2 - \Lambda)^{7/2}}
 G^{3,0} + \frac{9}{4} \Lambda^2 \frac{P}{(P^2 - \Lambda)^{5/2}}G^{2,1}\,,\\
 \dot{G}^{2,0} &=& -3\ \frac{P}{\sqrt{P^2-\Lambda}} G^{2,0}+ \frac{9}{4} \ \frac{\Lambda}{\left( P^2- \Lambda \right)^{3/2}} G^{3,0}\,, \\
 \dot{G}^{1,1}&=& -\frac{3}{2} \frac{\Lambda V}{(P^2- \Lambda)^{3/2}} G^{2,0}+\frac{9}{4} \frac{\Lambda PV}{(P^2- \Lambda )^{5/2}} G^{3,0} 
 -\frac{3}{4} \frac{\Lambda}{(P^2-\Lambda)^{3/2}} G^{2,1}\,, \\
\nonumber  \dot{G}^{0,2} &=& -3 \frac{\Lambda V}{(P^2- \Lambda)^{3/2}} G^{1,1}+3 \frac{P}{\sqrt{P^2-\Lambda}} G^{0,2}+ \frac{9}{4} \frac{\Lambda V P} {(P^2- \Lambda)^{5/2}} G^{3,0}- \frac{3}{4} \frac{\Lambda}{(P^2-\Lambda)^{3/2}} G^{2,1}\,.\\
 \end{eqnarray}
\end{widetext}

If we truncate the Hamiltonian (and hence the equations of motion) to
a given order, the equations of motion for quantum variables involve
higher order moments, as shown in the following equations to third
order in moments,
\begin{widetext}
\begin{eqnarray}
\dot{G}^{3,0} &=& -\frac{9}{2} \frac{P}{\sqrt{P^2 - \Lambda}} G^{3,0} + 
  \frac{9}{4} \frac{\Lambda}{(P^2 - \Lambda)^{3/2}} 
\left((G^{2,0})^2-G^{4,0}\right)\,,\\
\dot{G}^{2,1} &=& -\frac{3}{2} \Lambda 
   \frac{V}{(P^2 - \Lambda)^{3/2}} G^{3,0}+
  \frac{9}{4} \frac{\Lambda  PV}{(P^2 - \Lambda)^{5/2}} 
\left((G^{2,0})^2-G^{4,0}\right) + 
  \frac{3}{2} \frac{P}{\sqrt{P^2 - \Lambda}} G^{2,1}\,,\\
\dot{G}^{1,2} &=& -3 \Lambda 
   \frac{V}{(P^2 - \Lambda)^{3/2}} G^{2,1}-
  \frac{9}{2} \frac{\Lambda VP}{(P^2 - \Lambda)^{5/2}} 
\left(G^{1,1}G^{2,0}-G^{3,1}\right)+
  \frac{3}{2} \frac{P}{\sqrt{P^2 - \Lambda}} G^{2,1}\\
&&  - \frac{3}{4} \frac{\Lambda}{(P^2 - \Lambda)^{3/2}} 
\left(-4 (G^{1,1})^2 + G^{2,0}
G^{0,2}+3G^{2,2}\right)\,,\nonumber\\
\dot{G}^{0,3} &=& -\frac{9}{2} \Lambda 
   \frac{V}{(P^2 - \Lambda)^{3/2}} G^{1,2}-
  \frac{27}{4} \frac{\Lambda  VP}{(P^2 - \Lambda)^{5/2}} 
\left(G^{0,2}G^{2,0}-G^{2,2}\right)+
  \frac{9}{2} \frac{P}{\sqrt{P^2 - \Lambda}} G^{0,3}\nonumber\\
  &&+ \frac{9}{2} \frac{\Lambda}{(P^2 - \Lambda)^{3/2}} 
\left(G^{1,1}G^{0,2}-G^{1,3}\right)\,.
\end{eqnarray}
\end{widetext}
For consistency of the approximation one should eliminate in this
third-order approximation all terms of quartic order, i.e.\
fourth-order moments, producing a closed set of equations. (One may
eliminate terms quadratic in second-order moments as well, but they
are anyway subdominant when the $\hbar$ierarchy is satisfied.)
Truncating only the Hamiltonian to third order is thus not enough
because the general Poisson bracket between two third-order moments
produces terms of fourth order. (In general, the Poisson bracket of
two moments of order $n$ and $m$, respectively, produces terms of
order $n+m-2$ [see Eq. (\ref{GGbrackets})]. Only for second order does this result in a closed
Poisson algebra.) Consistently truncated equations of motion are
produced if one truncates the Hamiltonian as well as the Poisson
brackets to the order considered. An example to order five can be
found in the Appendix.

\subsection{High-order corrections}
\label{s:HighOrder}

Because of the complicated structure of the equations at higher
orders, analyzing them by analytical meanings is very hard. Hence we
need to resort to numerical methods. In particular, in this subsection
we will numerically solve the evolution equations, derived using
computer algebra, up to 10th order.  An important question is then
what values to choose for all the 63 moments involved at initial
time $\phi_0=0$. Here we
will simply choose the initial state corresponding to an unsqueezed
Gaussian pulse in the volume,
\begin{equation}
\Psi(\chi)=\frac{1}{\pi^{1/4}\sqrt{\sigma}}\,e^{-\frac{(\chi-V_0)^2}{2\sigma^2}
+\frac{i P_0 \chi}{\hbar}}.
\end{equation}
While this state is very special in the space of all semiclassical moments,
it already serves well to demonstrate several features of quantum
evolution. A more detailed analysis of the large parameter space will
appear elsewhere.

Using (\ref{defG}), the initial moments are then given by,
\begin{eqnarray}
G^{a,b}&\equiv&\sum_{d=0}^{{\rm Min}(a,b)}(-i\hbar)^a C_{ab}^d \int_{-\infty}^{\infty}\md \chi
\,\Psi^*(\chi) (\chi-V_0)^{b-d}
\nonumber\\\label{integral}
&&\qquad\qquad\times\left(\frac{\md}{\md \chi}-i\frac{P_0}{\hbar} \right)^{a-d}\,\Psi(\chi),
\end{eqnarray}
where ${}^*$ denotes complex conjugation.
This integral can be performed and happens to be vanishing except in those
cases for which both $a$ and $b$ are even numbers,
\begin{eqnarray}
G^{a,b}=
\left\{
\begin{array}{c}
2^{-(a+b)} \hbar^a\sigma^{b-a}\frac{a!\,b!}{\left(\frac{a}{2}\right)!\left(\frac{b}{2}\right)!}
\quad{\rm if}\, a \,{\rm and}\, b\, {\rm are}\,{\rm even},\\
\\
 0\quad\quad\quad\quad\quad\quad{\rm otherwise}.
\end{array}
\right.
\end{eqnarray}
This state saturates the uncertainty relation, and we have that $\sigma\sim
\Delta V=\sqrt{G^{2,0}}\sim O(\sqrt{\hbar})$, so the moments satisfy
the semiclassical $\hbar$ierarchy. In our simulations, however, we
have used $\hbar=1$ in order to make implications for the behavior of
back-reaction more clearly visible. Nevertheless, the features we
point out have been verified also for smaller values of $\hbar$.


By this choice of initial values, the many-parameter family of all
moments up to a given order is thus reduced to a 1-parameter family
labeled by the volume spread $\sigma$. From this perspective,
choosing a Gaussian initial state seems rather special and arbitrary;
it is distinguished only by the simplicity of writing down a wave
function. As the system evolves, the state will depart from Gaussian
behavior, a property which can easily be seen by following the
behavior of the moments. Before discussing the numerical results in
this direction, we note that the choice of Gaussian moments as initial
values can sometimes be dangerous also for numerical purposes. A
Gaussian state saturates the uncertainty relations, and numerical
errors may easily move the evolved moments to the wrong side of
saturation. Depending on the dynamical system, the saturation surface
may be unstable such that uncertainty relations may be violated after
some stretch of numerical evolution, which at that stage could no
longer be trusted. (See also \cite{HighDens} for a discussion of
similar issues in small-volume regimes.) In the numerical evolutions
shown here, the fact that uncertainty relations are maintained has
explicitly been tested by monitoring the relevant combinations of the
second-order moments.  In particular, the final time of each evolution
has been chosen as the last time where the convergence tests are
obeyed.

Once an unsqueezed Gaussian is chosen as initial state, the only
choice left is the value of $\sigma$. The system under consideration
has no ground state, and thus there is no preferred value for $\sigma$
as for instance the harmonic oscillator would suggest. For small
initial fluctuations, which are related to $\sigma$ by $\sigma\sim
\Delta V\sim 1/\Delta P$, we just require that $1/P_0\ll\sigma\ll V_0$
is satisfied for the initial values. Specifically, in all evolutions
presented in this paper we take $V_0=1$, $P_0=10^4$ and $\sigma=0.01$.
This does not represent a high restriction since we have also
performed numerical simulations for different larger values of the
width ($\sigma=1, 100, 10^4$) and the qualitative picture is not
changed.  Even so, the behavior of the system for values of
$\sigma\lesssim 10^{-4}$ is completely different. In this case the
corrections to the classical trajectories are very large from the very
beginning and the approximation can not be regarded as valid. This
happens because in the evolution equations (\ref{dotV}) and
(\ref{dotP}), the moments $G^{n-1,1}$ and $G^{n,0}$ appear.  At the
initial time $\phi_0$ the former is zero, whereas the latter is of the
order $\sigma^{-n}$ for even $n$ and vanishing for odd $n$.  Hence,
for very small $\sigma$, the initial time derivatives of $V$ and $P$
increase much with the considered order. This entails a completely
different trajectory for the classical objects at each order and thus
the approximation breaks down.

\begin{figure}[ht]
\includegraphics[width=0.5\textwidth]{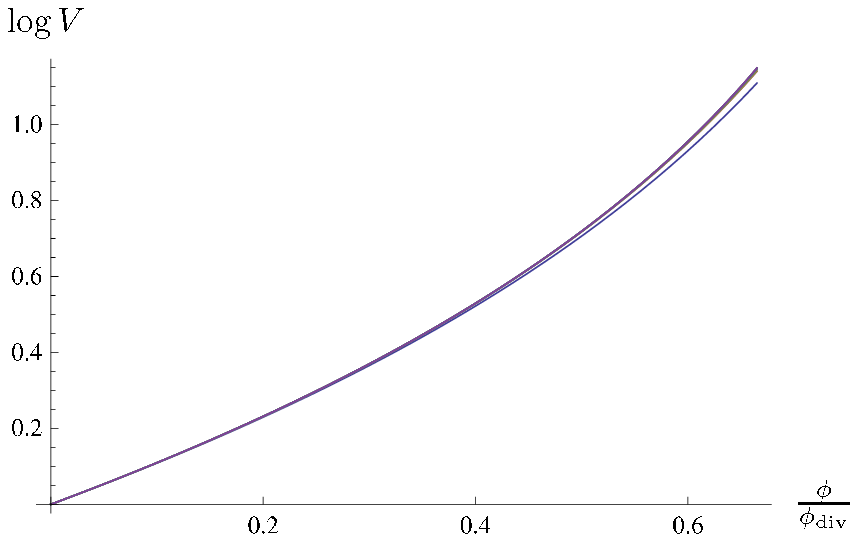}
\includegraphics[width=0.5\textwidth]{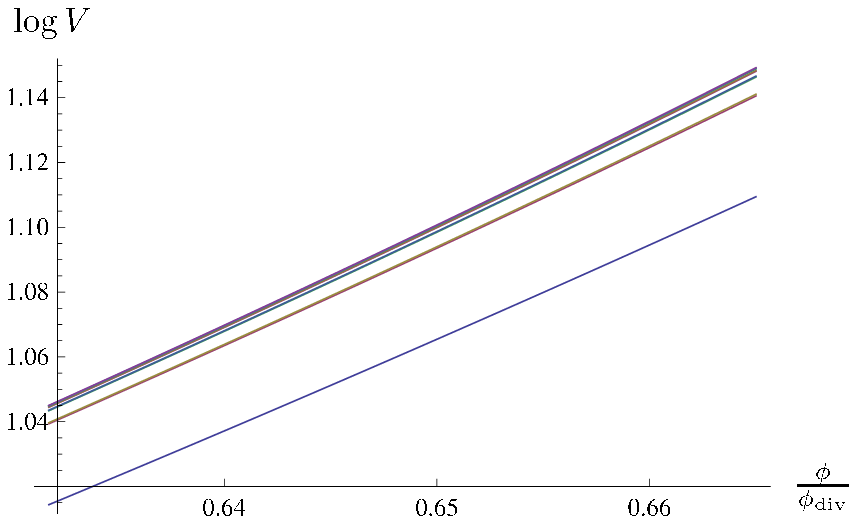}
\caption{\small The ten different evolutions of the volume $V$ at each
order for $\Lambda=9\times 10^7$.  The lowest trajectory corresponds
to the classical one. The values are increasing with the considered
order in such a way that the largest trajectory is obtained at 10th
order. For a more detailed analysis of the convergence, see
App.~\ref{convergence}.}
\label{Vplot}
\end{figure}

Finally, for the classical solutions (\ref{Pclass}), (\ref{Vclass}) to
exist, the cosmological constant must be in the interval $[0,P_0^2)$,
so we have chosen three representative values: small $\Lambda=1$,
intermediate $\Lambda=10^4$, and large $\Lambda=9\times 10^7$. These
values are much higher than the observed one. Even so, since the model
we adopt as an example is unphysical anyway, it would not be
meaningful to have it clad in a physical guise by using the
observational value for $\Lambda$. We use the mentioned values in
order to bring out more clearly the properties we have found about
high-order moments.


\subsubsection{Numerical results}

\begin{figure}
\includegraphics[width=0.5\textwidth]{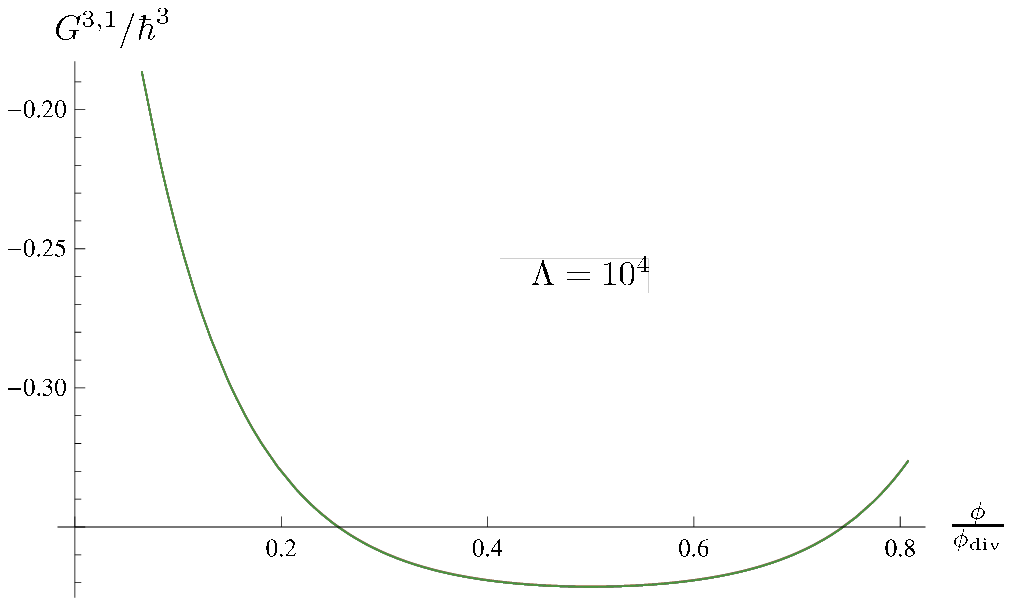}
\includegraphics[width=0.5\textwidth]{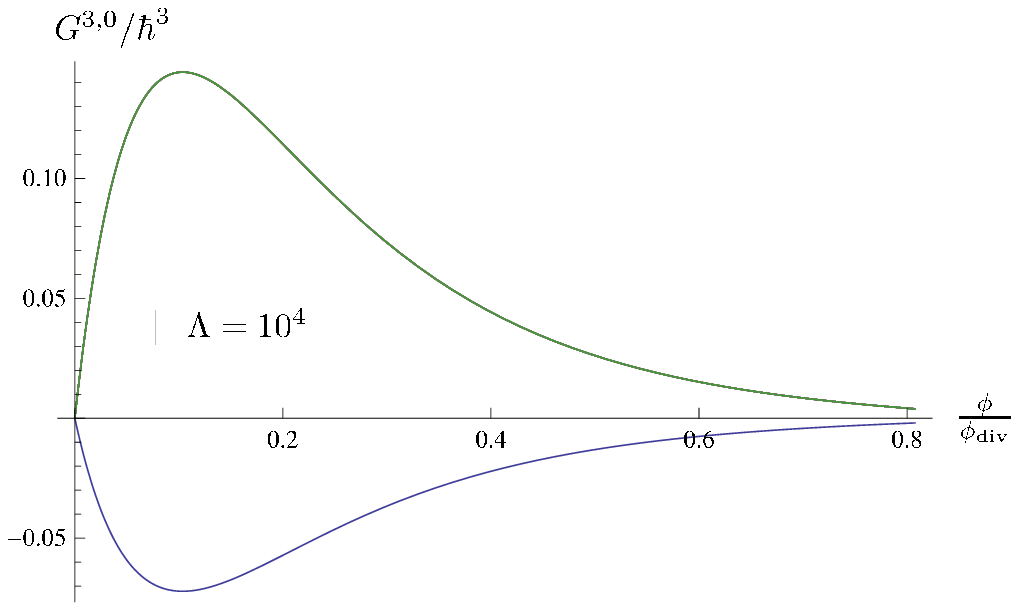}
\includegraphics[width=0.5\textwidth]{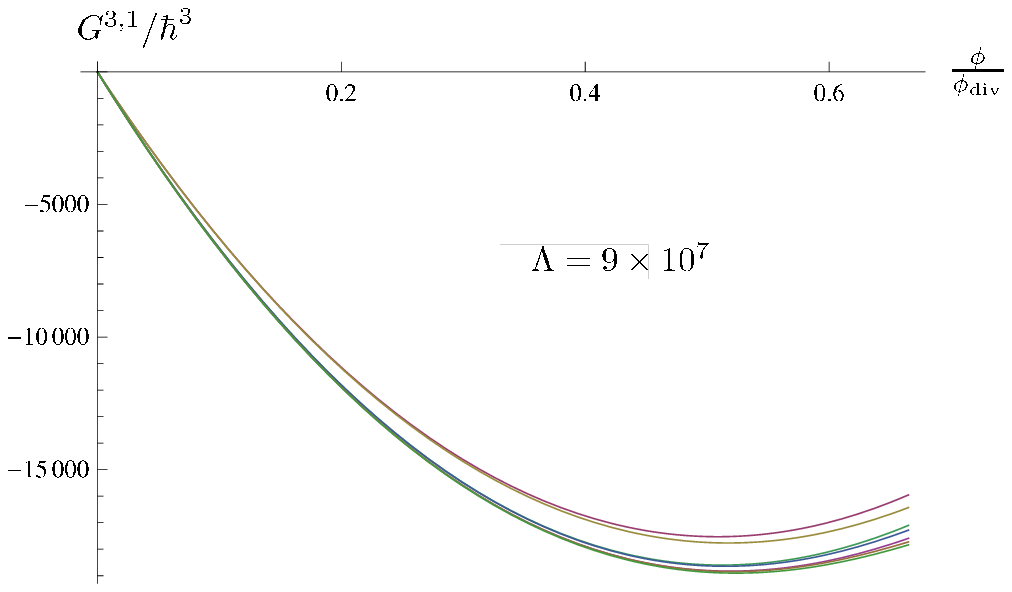}
\includegraphics[width=0.5\textwidth]{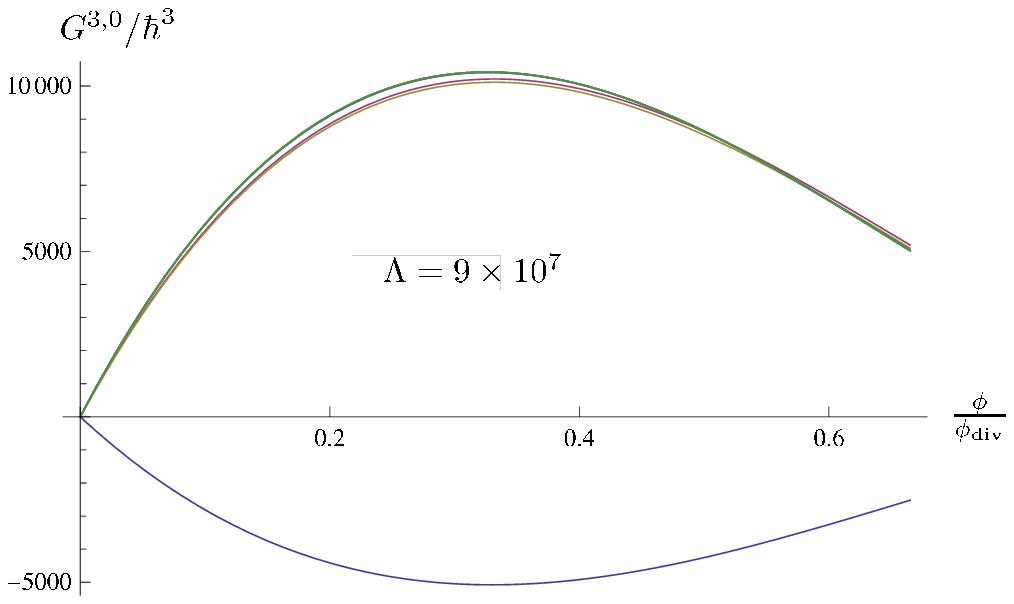}
\caption{\small An example with two different moments with one and two
branches respectively.  For every moment $G^{i,j}$, the evolution at
every order from $i+j$ to 10 is shown.  The first two plots
correspond to $\Lambda=10^4$, whereas the other two
represent the evolution of the same moments for
$\Lambda=9\times10^7$.}
\label{moments}
\end{figure}

Let us first explain the behavior of the volume $V$. For the case of
small and intermediate values of the cosmological constant, the
evolution of the volume reproduces the classical one with small
corrections at all orders.  However, in the case of large cosmological
constant we find that, at later times, the classical trajectory of the
volume receives large quantum corrections, which make it diverge
faster (see Fig. \ref{Vplot}). In particular, we do not see any
indication that quantum back-reaction may trigger a recollapse;
instead, the classical divergence of the volume is enhanced by quantum
corrections with each order. In App. \ref{convergence} we provide a
complete analysis of the convergence of our solutions at different
orders by studying the relative corrections that the expectation value
of the volume receives at each order. These results show that our
effective analysis is valid (converges exponentially) at all considered
times with a high precision.

Regarding the evolution of the moments there are two different
behaviors. On the one hand, those $G^{i,j}$ with $(-1)^{i+j}=1$ only
show one branch; see the two characteristic examples shown in
Fig. \ref{moments}. That is, their evolution at different orders
reproduce the same trajectory (for large values of the cosmological
constant with small corrections). On the other hand, the moments with
$(-1)^{i+j}=-1$ show two different branches. One of them always
corresponds to the order $i+j$ and the other one to higher orders. This
curious behavior is due to the fact that, in the latter case, when
truncating the equations of motion corresponding to this moment at
order $i+j$, we neglect the contribution of certain moments that are
initially nonvanishing. In the former case, for $G^{i,j}$ with
$(-1)^{i+j}=1$, all neglected moments in its equation of motion at
order $i+j$ happen to be initially vanishing.

\begin{figure}[ht]
\includegraphics[width=0.45\textwidth]{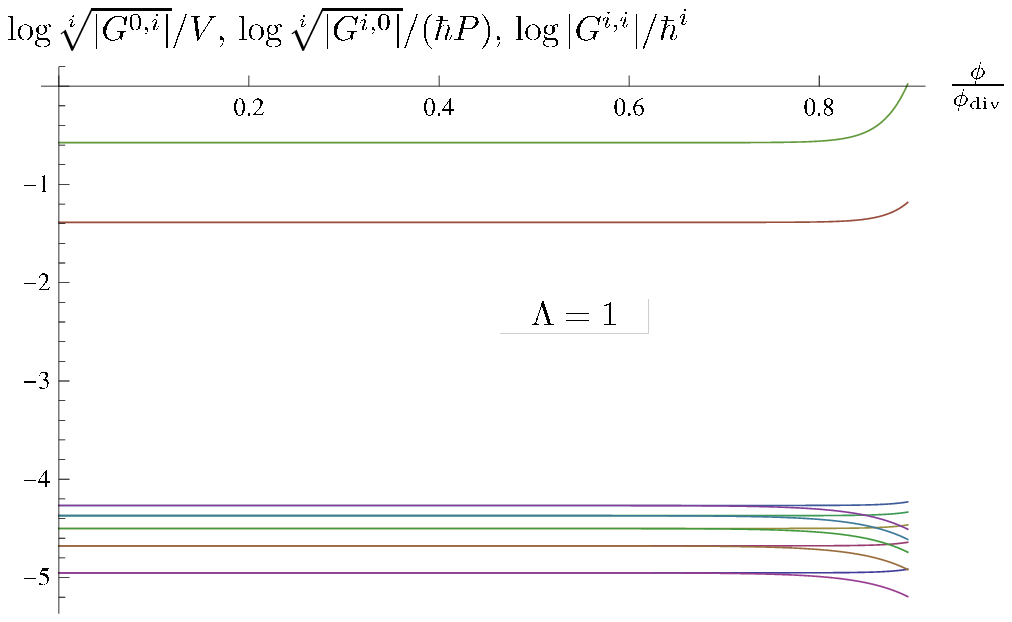}
\includegraphics[width=0.45\textwidth]{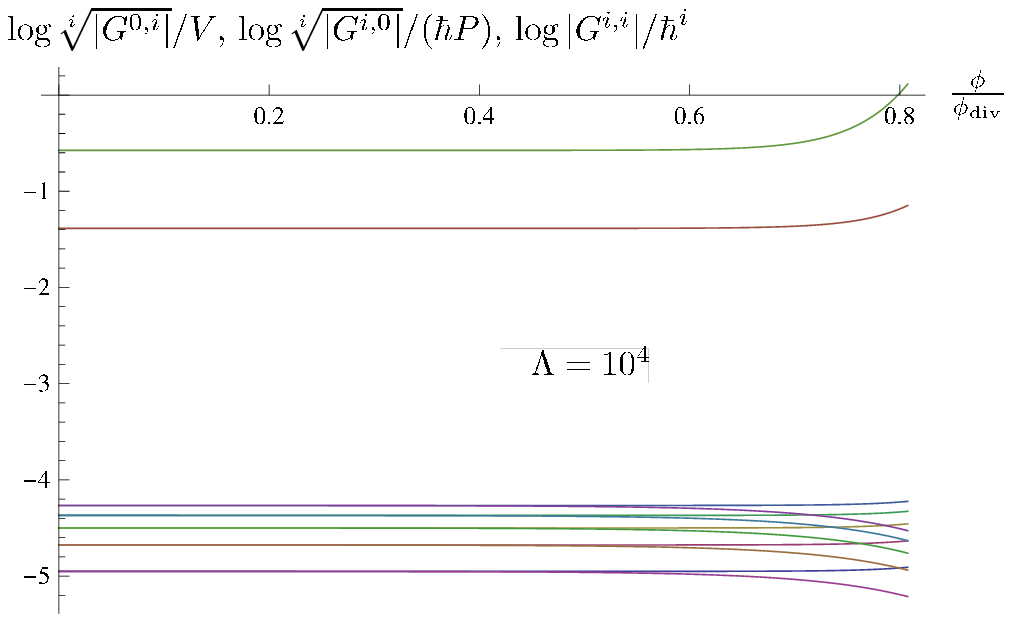}
\includegraphics[width=0.45\textwidth]{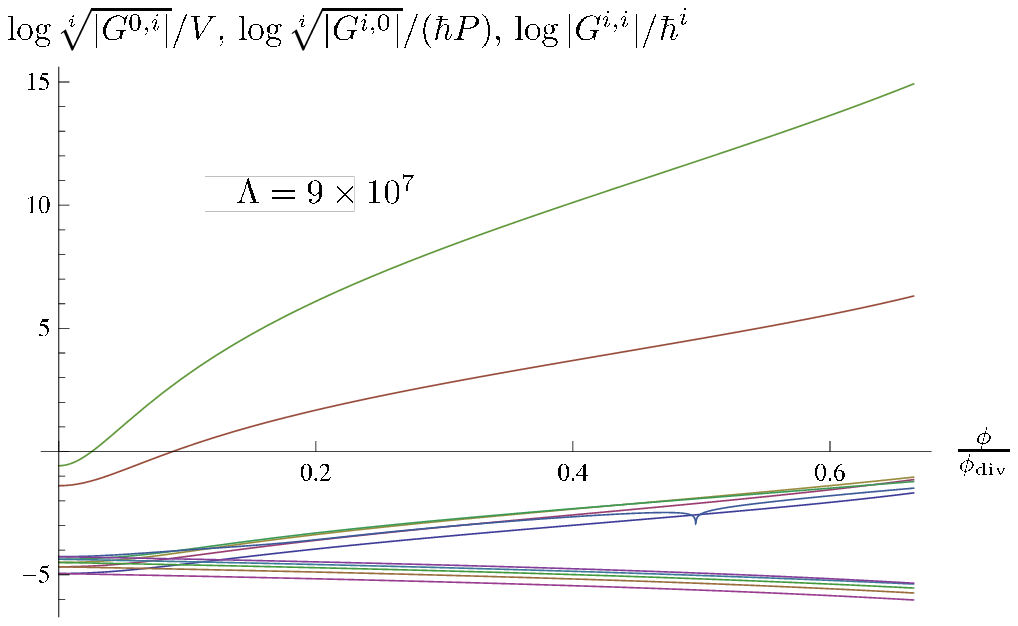}
\caption{\small These figures show, in a logarithmic scale, objects
that are constant in the case of a vanishing $\Lambda$.  The plots
correspond to the values $\Lambda=1, 10^4,$ and
$9\times 10^7$ respectively.  In the first two graphics the moments
are constant throughout evolution until very late times. On the other
hand, for a large cosmological constant this behavior is violated from
the very beginning. As a general feature, in all cases, once the
moments are no longer constant, the quantities $\sqrt[i]{|G^{0,i}|}/V$ as
well as $|G^{i,i}|$ happen to be increasing functions, whereas
$\sqrt[i]{|G^{i,0}|}/P$ are decreasing. The different lines
correspond, from largest to smallest value at the initial time, to the
moments $G^{4,4}$, $G^{2,2}$, and then, with an equal initial value by
pairs, to $(G^{10,0}, G^{0,10})$,
$(G^{8,0}, G^{0,8})$, $(G^{6,0}, G^{0,6})$, $(G^{4,0}, G^{0,4})$, and $(G^{2,0}, G^{0,2})$.}
\label{fig1}
\end{figure}

\begin{figure}
\includegraphics[width=0.45\textwidth]{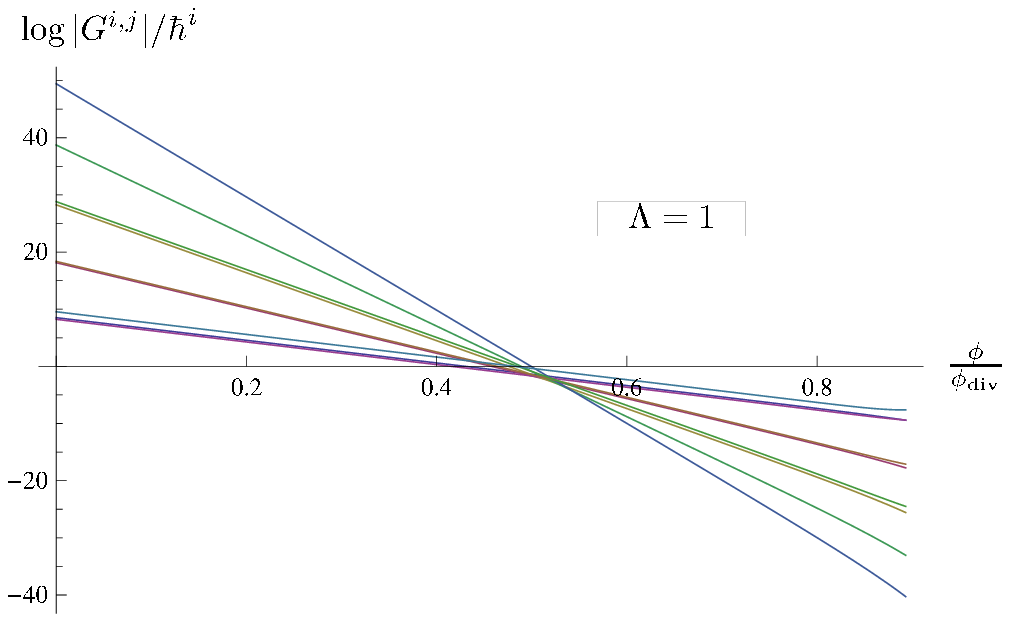}
\includegraphics[width=0.45\textwidth]{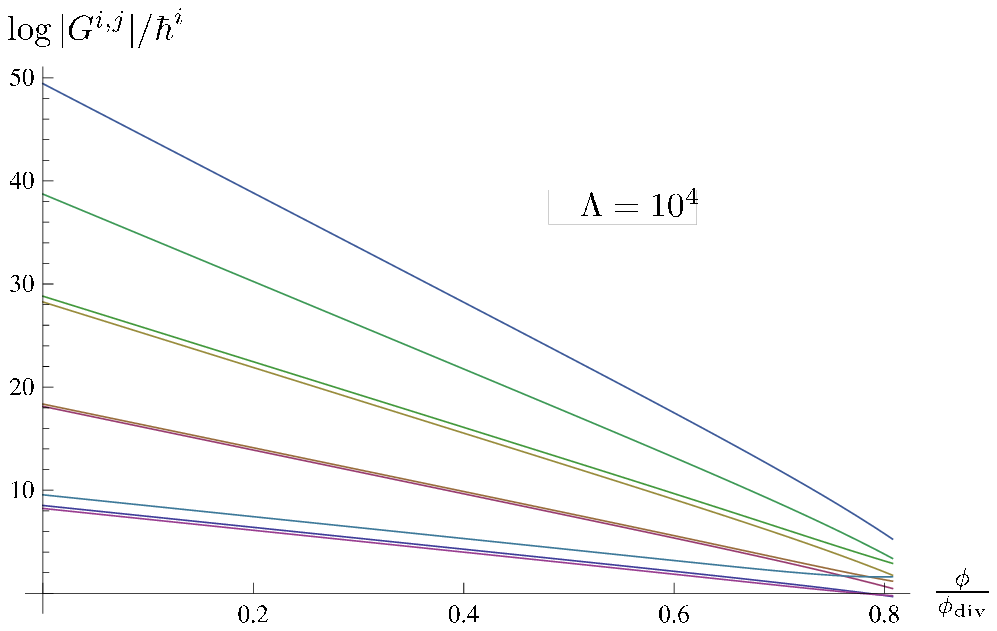}
\includegraphics[width=0.45\textwidth]{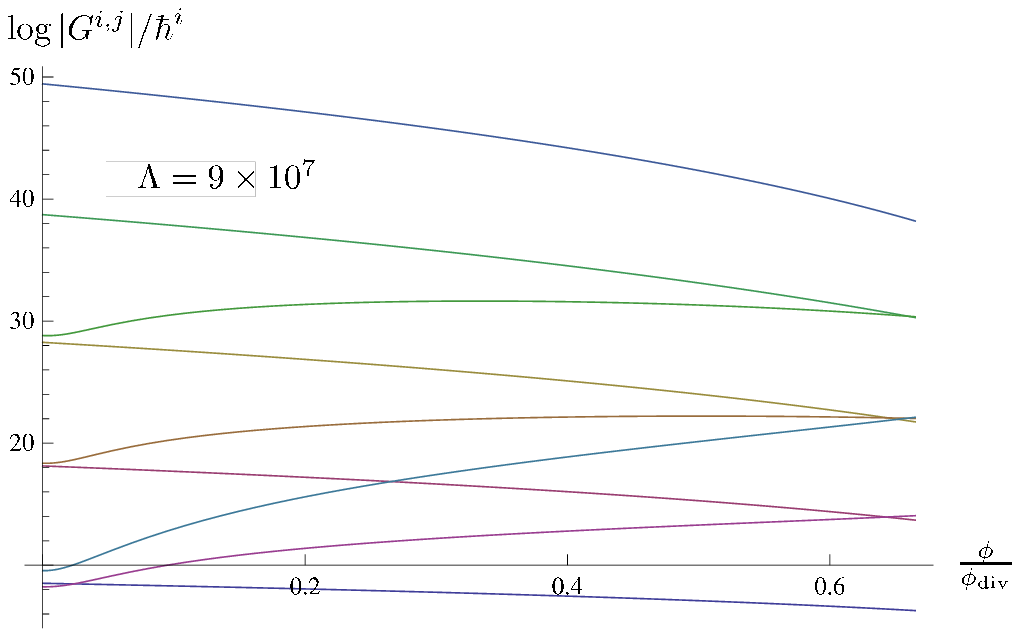}
\caption{\small This figure shows the evolution of the moments
$G^{i,j}$ with $i>j$ and for both $i$ and $j$ even. Different plots
correspond to the values $\Lambda=1, 10^4,$ and $9\times 10^7$
respectively. In the first two plots we observe that the exponentially
decaying behavior of these moments is maintained for small and
intermediate values of the cosmological constant.  On the other hand,
for large cosmological constant, the exponential decay is not followed
anymore and there are even some moments that slightly
increase. Interestingly, at the final time of the evolutions shown
here, different moments $G^{i,j}$ with the same index $i$ approach
a common value. From largest to smallest values at the accumulation time
the presented moments are: $G^{10,0}$ alone; $G^{8,2}$ with $G^{8,0}$;
$G^{6,4}$ with $G^{6,2}$ and $G^{6,0}$; $G^{4,2}$ with $G^{4,0}$; and
finally $G^{2,0}$ on its own. The tendency to this behavior can
already be noted in the second plot, whereas in the first one all
moments gather at a value near $1$.}
\label{fig2}
\end{figure}

\begin{figure}
\includegraphics[width=0.45\textwidth]{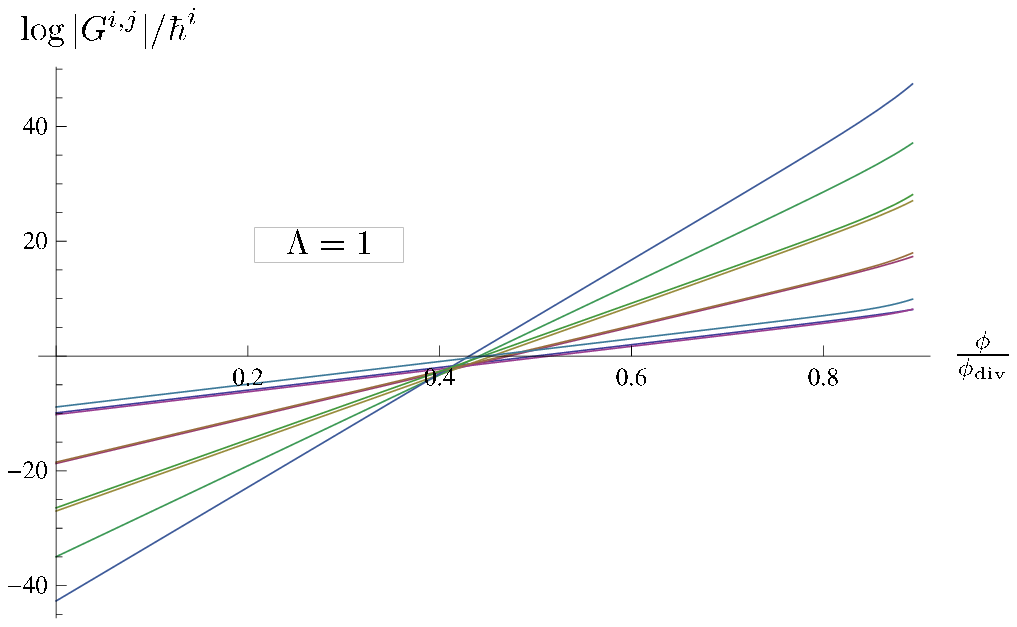}
\includegraphics[width=0.45\textwidth]{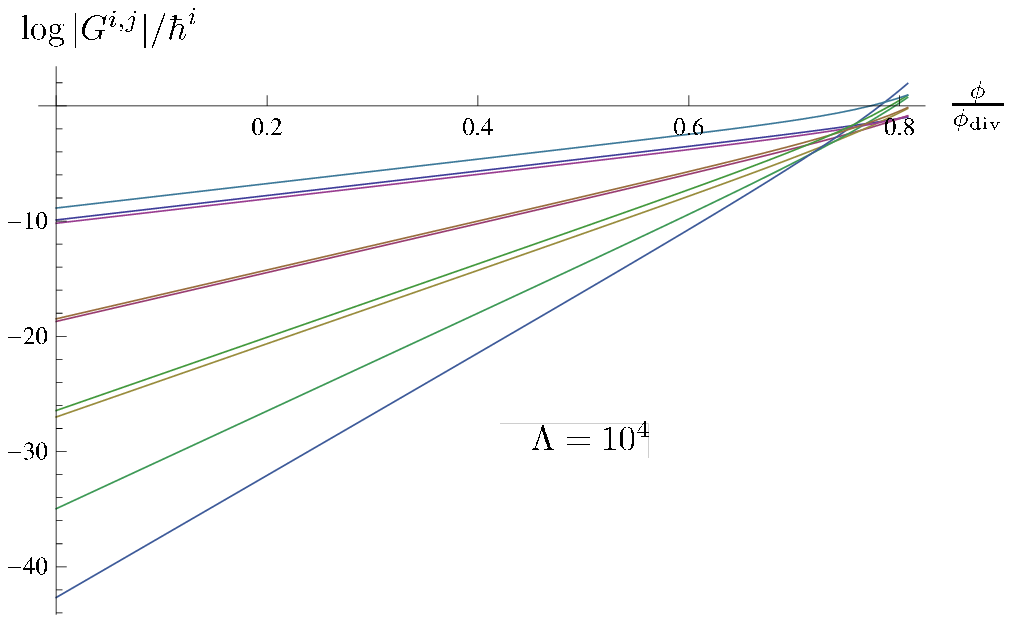}
\includegraphics[width=0.45\textwidth]{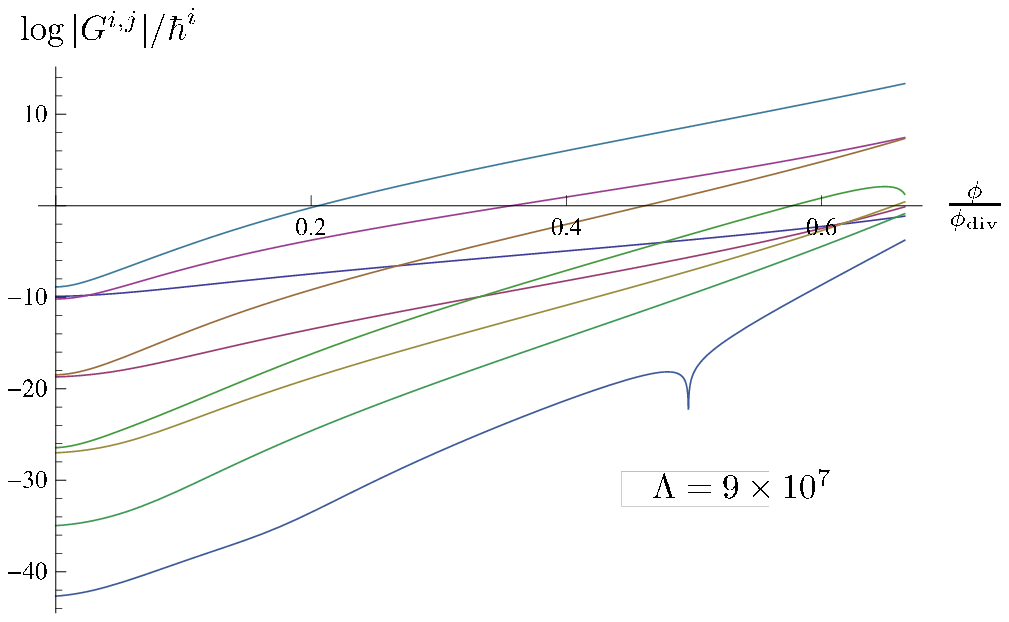}
\caption{\small This figure corresponds to $G^{i,j}$ with $i<j$ and
both $i$ and $j$ even numbers.  The value of the cosmological constant
is again $\Lambda= 1, 10^4$ and $9\times 10^7$ for different plots. As
in the previous case, in the first two plots all moments coincide at a
value near 1 at a given time, whereas in the last plot the different
variables tend to gather in small sets given by those moments with the
same number of $\hat P$. In this last plot, at a final time and from
largest to smallest value: $G^{4,6}$, then the three $G^{2,i}$, and
finally the five $G^{0,i}$. Spikes in this and the following plots
arise from transitions of the moments through zero, shown in
logarithmic form.}
\label{fig3}
\end{figure}

\begin{figure}
\includegraphics[width=0.45\textwidth]{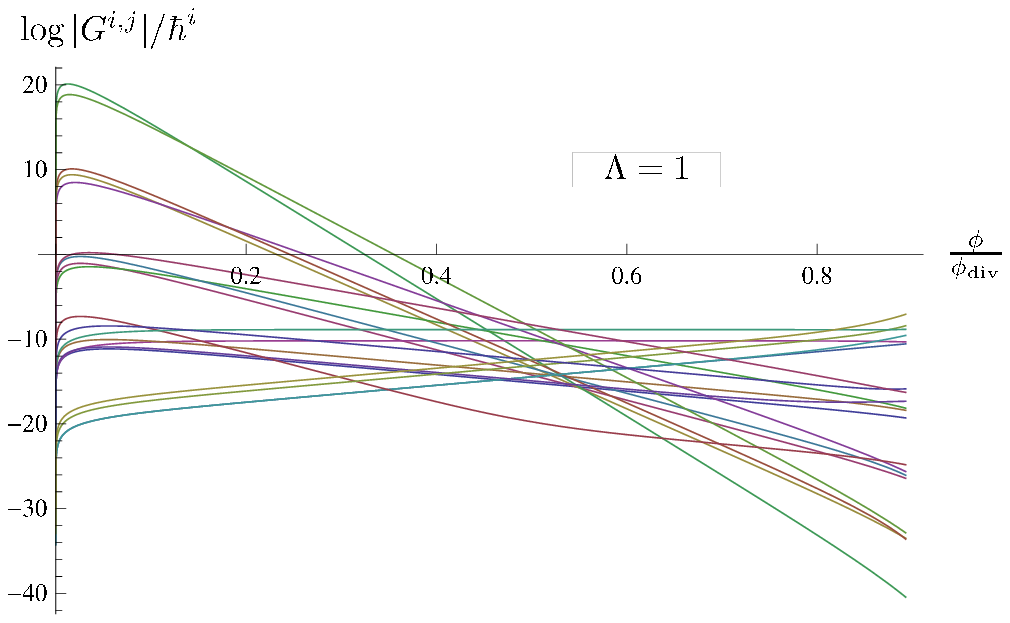}
\includegraphics[width=0.45\textwidth]{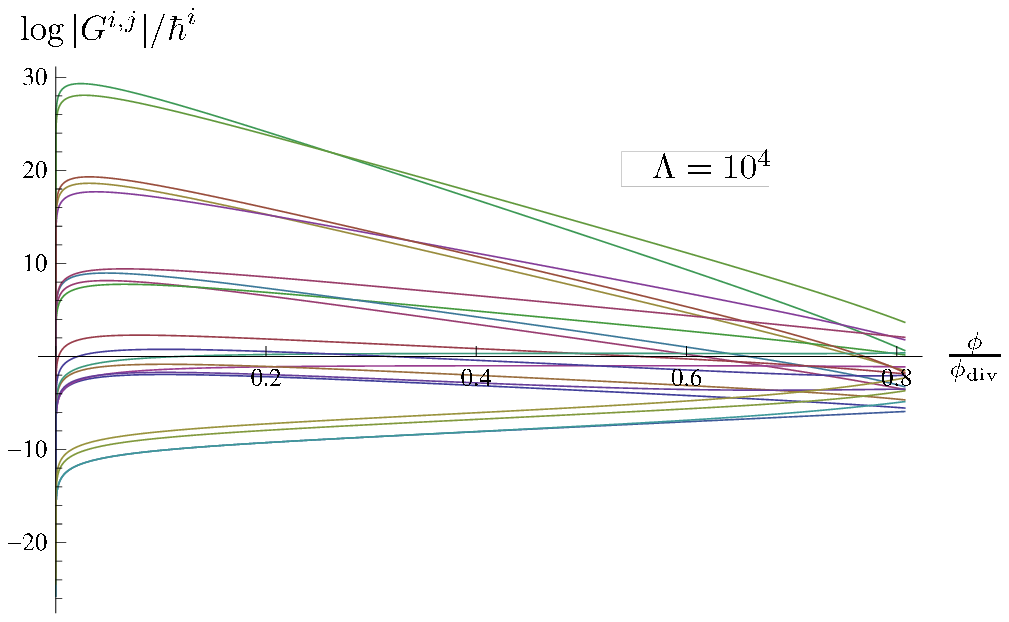}
\includegraphics[width=0.45\textwidth]{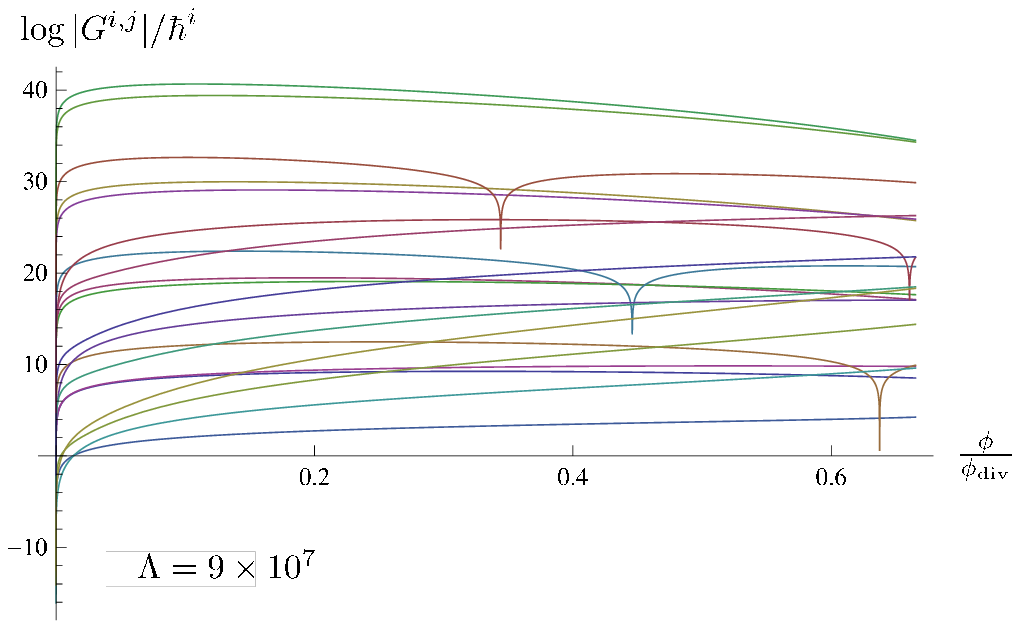}
\caption{\small This figure corresponds to initially vanishing
$G^{i,j}$ with $i>j$.  The same values of $\Lambda$ as in previous
cases for each plot apply.  Since the initial state we have picked is
not adapted to the equations of motion, at the initial time all these
modes are excited. Their absolute value after this excitation is
approximately ordered by increasing $(i-j)$. That is, the largest
moment is $G^{9,0}$, then $G^{9,1}$, $G^{8,1}$, \ldots On the other hand,
it is also interesting to note that this excitation becomes larger
with an increasing $\Lambda$. Finally, in the case with vanishing
$\Lambda$, all these moments are exponentially decreasing. This
behavior is qualitatively disappearing as one considers a larger
cosmological constant.}
\label{fig4}
\end{figure}

\begin{figure}[h!]
\includegraphics[width=0.5\textwidth]{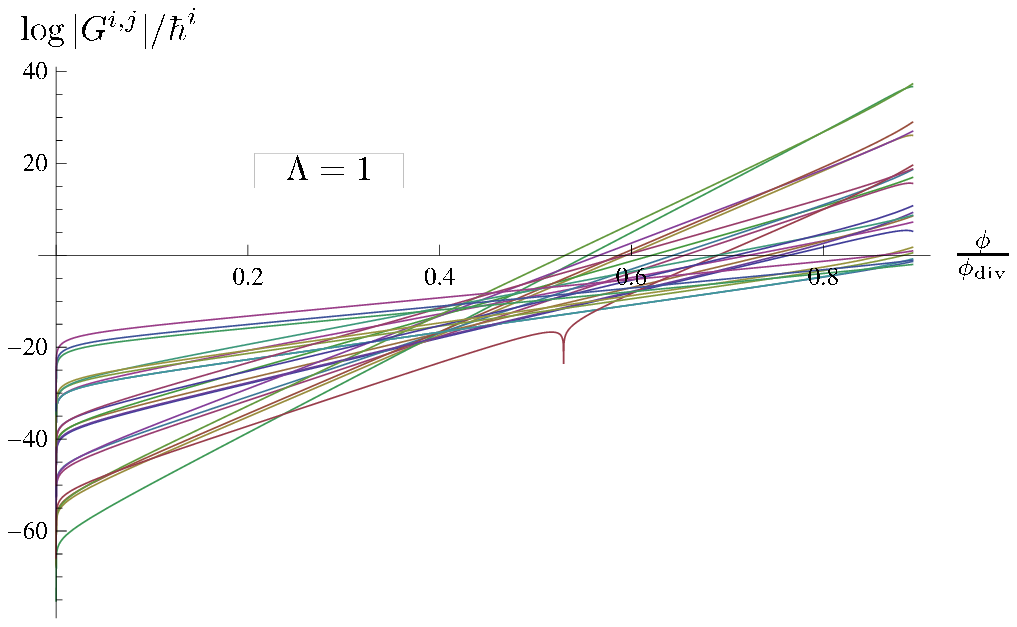}
\includegraphics[width=0.5\textwidth]{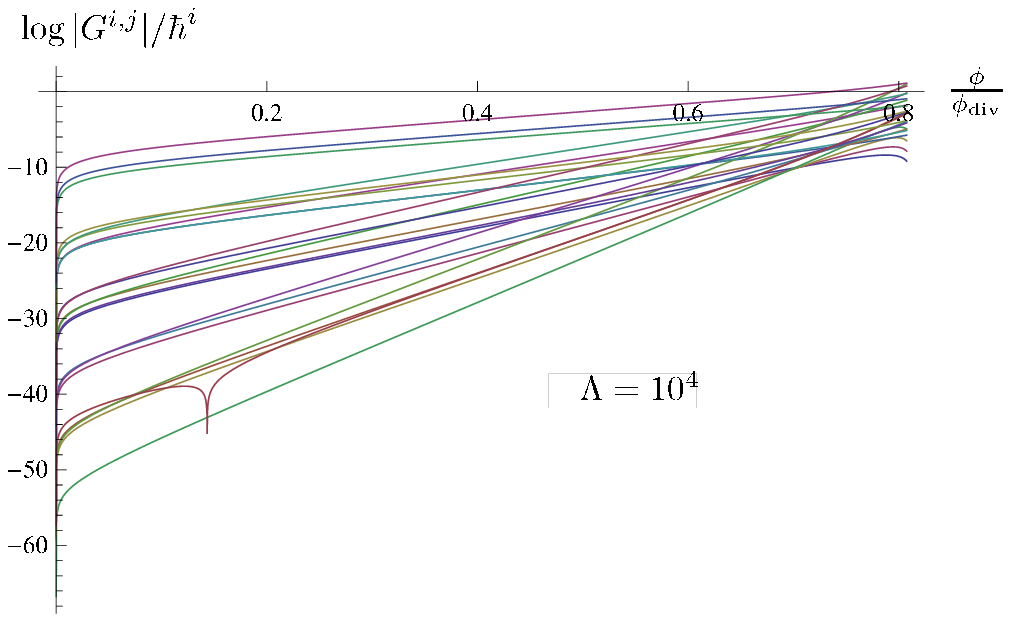}
\includegraphics[width=0.5\textwidth]{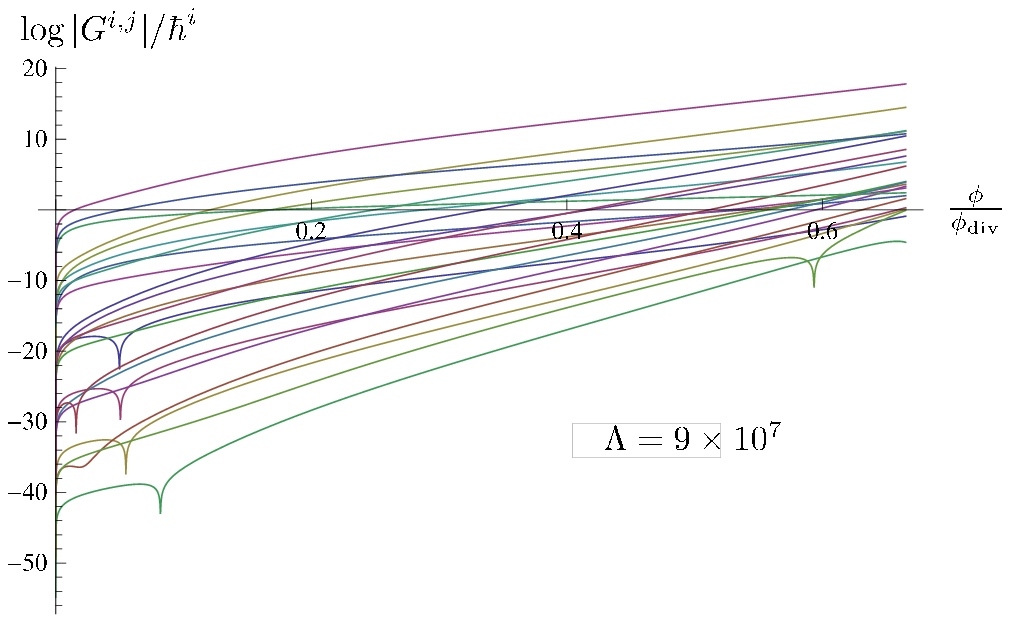}
\caption{\small This figure shows the initially vanishing moments
$G^{i,j}$ with $i\leq j$.  The same values of $\Lambda$ as in previous
cases for each plot apply. As in the previous case, initially all
moments are excited to a given value, which is larger with an
increasing $(i-j)$. In this way, the moments with lowest absolute
value are $G^{0,9}$, $G^{1,9}$, $G^{1,8}$, \ldots On the other hand,
and contrary to the previous case, the qualitative behavior of these
moments is kept the same as in the case with $\Lambda=0$ even for
large values of the cosmological constant.}
\label{fig5}
\end{figure}

Before commenting on the collective behavior of different moments, it
will be useful to remember the particular case with $\Lambda=0$ where,
as shown in Sec. \ref{L0},
analytic solutions for the moments at all orders can be obtained. In
this case the evolution equations for the classical variables as well
as for the quantum corrections are completely decoupled, and there is
no quantum back-reaction.  Using the exact solutions (\ref{GabSol}),
the moments $G^{ab}$ with $a>b$ are exponentially decreasing,
whereas those with $a<b$ increase exponentially. On the other hand,
the variables with $a=b$ are constants of motion. Note also that the
combinations $\sqrt[i]{|G^{0,i}|}/V$ and $\sqrt[i]{|G^{i,0}|}/P$ remain
constant under evolution. In Fig. \ref{fig1} we show these objects
for different values of the cosmological constant. It can be clearly
seen how the trajectories depart from the constant behavior earlier
when increasing the value of the cosmological constant. In this simple
system one can easily construct many other constants of motion, e.g.
$G^{a,b}/(P^aV^b)$. Even though, for illustration, we will only consider
those mentioned above.

Finally, in Fig. \ref{fig2} to \ref{fig5} we show the collective
behavior of the different moments $G^{i,j}$.  Essentially they have
been divided in four groups depending on whether both $i$ and $j$ are
even numbers or not and whether $i<j$ or $i>j$. This classification is
inherited on the one hand from our chosen initial conditions, where
the only nonvanishing moments are those with both $i$ and $j$ even,
and on the other hand from the case without cosmological constant
explained above, where the increasing or decreasing behavior of
different moments also depends on the sign of $i-j$.

The increasing behavior of those moments $G^{i,j}$ with $i<j$ is
obeyed also for nonvanishing values of the cosmological constant (see
Fig.  \ref{fig3} and \ref{fig5}). On the other hand, in
Fig. \ref{fig2} and \ref{fig4} the decreasing moments are plotted, but
this behavior is violated as we move to higher values of the
cosmological constant. As a general feature, in these plots it can
also be observed that the evolution of different moments seems to be
``slowed down'', that is, everything (for example the instant when
nearly all moments cross each other) happens latter as one increases
the value of the cosmological constant.  This makes the hierarchy
between different moments be retained for a longer time.

Another interesting issue is shown in the plots corresponding to the
initially nonvanishing moments (Fig. \ref{fig2} and \ref{fig3}): at
late times those moments $G^{i,j}$ with the same number of $\hat P$,
that is with a common index $i$, accumulate at a common value.

Finally, we note how the initially vanishing moments behave at the
beginning of the simulations. As can be seen in Fig. \ref{fig4} and
\ref{fig5}, since the initial state of the system is not adapted to
the equations of motion, these moments are excited very quickly to a
``natural'' value which more closely corresponds to a dynamical
coherent state. The absolute value of a given moment $G^{i,j}$
immediately after this excitation, increases with the value of the
cosmological constant $\Lambda$ and approximately also with the
difference $i-j$.  This last dependence can be seen in the equations
of motion.  More explicitly, making use of the general formula
(\ref{GGbrackets}) and our particular Hamiltonian
(\ref{ourhamiltonian}) it can be checked that the time derivative of a
moment $G^{i,j}$ is given as a linear combination of the following
objects $\{G^{i,j-1}G^{k-1,0}, G^{i-1,j}G^{k-1,0}, G^{i,j-1}G^{k-2,1}$
$,G^{i+k-n,j-n}, G^{i+k-1-n,j+1-n}\}$, where $k$ must be summed from 2
to the order $K$ at which we decide to truncate the Hamiltonian, and $n$
corresponds to the sum over odd numbers that appears in
Eq. (\ref{GGbrackets}).  Taking into account the dependence of these
objects on the Gaussian width $\sigma$ at the initial time, it is
straightforward to see that the initial time derivative of $G^{i,j}$
behaves like $\sigma^{-(K+i-j)}$.

\subsubsection{Implications for state evolution}

In Sec. \ref{2ndorder},
several questions about the evolution of states has been suggested by
the second-order analysis, but could not be answered without more
general information about the high-order system of quantum
back-reaction. With the numerical results, we can now provide
additional indications regarding quantum evolution, but also new
properties become visible.

First, the strong increase of some of the moments already seen at
second order is confirmed. The system is thus necessarily one of
strong quantum back-reaction. Results about self-adjoint extensions of
quantum Hamiltonians have suggested that the evolution can be
continued through the classical divergence of the volume. The only
intuitive semiclassical interpretation would be that quantum
corrections become so strong that they can trigger a recollapse. Our
numerical results to the orders specified do not provide any
indication for this; they rather show that the divergence is enhanced
by quantum corrections.

The behavior of the moments confirms two expectations about state
evolution. First, several moments grow and become dominant providing
strong quantum back-reaction. Secondly, the state rapidly departs from
the initially chosen Gaussian moments. While it is difficult to
reconstruct a wave function from the moments, it is clear from several
of their properties that the state cannot remain Gaussian. A Gaussian
state has vanishing moments of odd order, a property which is
immediately violated once the state is evolved. As the rapid initial
increase of the odd-order moments shown in Figs.~\ref{fig4} and
\ref{fig5} shows, the evolution quickly adapts the state to one whose
moments change less severely. To some degree, even the special choice
of a Gaussian state may not be much of a restriction since the
evolution soon leads to a more suitable dynamical state. However, for
robust conclusions about quantum evolution from a fixed set of initial
values one must analyze how the state settled down to
depends on the initial moments. If different sets of initial moments
still lead to adapted states, but ones that differ from the Gaussian
adapted state in a way sensitive to the initial set, the quantum
behavior would be hard to predict without knowledge of what state may
be preferred. Here, much more numerical analysis of the large
parameter space involved is necessary.

Finally, we see from figures such as Figs.~\ref{fig4} and \ref{fig5}
that a hierarchy of the moments is maintained for a rather long time
even throughout the phase of dynamically adapting the state. Moments
of different orders clearly fall into distinct classes as for their
behavior of decay or increase. Only when moments of different orders
converge, which interestingly happens in a very narrow time frame for
all orders shown, will the hierarchy be broken. At such a point, the
state can no longer be considered semiclassical, and truncations
of the infinite system of effective equations (and the asymptotic
expansion they represent) become more difficult to justify in general
terms. Nevertheless, in our specific case, even when the state is no
longer of recognizably semiclassical form, the truncation is still
justified by the convergence analysis we perform in
App.~\ref{convergence}. We have also shown the evolution even after
that time of accumulation because it indicates another feature. After
the accumulation, the moments again separate into clearly demarcated
sets, indicating that another hierarchy arises.  However, since one
has to evolve through an anhierarchical point, it is not clear whether
this feature is one of the full system or an artifact of the
truncation.  Finally and somewhat unexpectedly, the hierarchy is
maintained longer for larger values of the cosmological constant, even
though the deviation from the harmonic model is then stronger.

\section{Discussion}

The main contribution of this article is the introduction of a new
computational method to analyze quantum back-reaction of quantum
mechanical or quantum cosmological systems. We have
shown that the use of efficient computer algebra tools in combination
with the closed formula for the Poisson brackets of two generic
moments has been essential to push the feasibility of computations to
very high order. In particular, the example we have studied
of a spatially flat, isotropic universe with a positive cosmological
constant and a free, massless scalar field already indicates the usefulness
of these methods.

Our analysis has found several new properties of
state evolution, some of which were quite unexpected. For example,
the state rapidly deviates from the initial Gaussian form (in the volume), but then
settles down to another shape obeying a hierarchy of the moments. In
this range, truncations used to analyze effective equations remain
justified. At some point, the moments converge, interestingly at about
the same time. Several properties found here remain without an
analytical explanation, stimulating further studies. For instance,
somewhat counterintuitively, the moments happen to
``slow down'' their evolution as the value of the cosmological
constant increases. That is, qualitatively they follow the same
pattern, independently of $\Lambda$,
but everything happens later for large values of the cosmological
constant. Another interesting feature comes from the evolution
of the initially non-vanishing moments $G^{i,j}$: at late times they
accumulate at a common value for each index $i$.

On the other hand, we have also studied the convergence of this
truncated system of equations with respect to the order (see
App. \ref{convergence}) by analyzing the relative error in the
trajectory of the expectation value of the volume $V$.  Remarkably,
even though a priori we expect only asymptotic rather than convergent
expansions, the results converge exponentially for all considered
times within a large range of orders. This means that our results
reliably reproduce the full quantum behavior of the system even quite
near the divergence.

Finally, details of the numerics remain to be explored, most
importantly those related to the large parameter space involved. For
instance, it is not clear yet how strong the role of the choice of an
initial state is. The shape of the state changes rapidly in a very
brief initial phase, as shown by a large change in the moments, and
then settles down to a form better conserved by the dynamics. This
evolved state seems adapted to the dynamics, but it is not known at
present whether differently chosen initial states will give rise to
the same kind of dynamically evolving state, nor is it known whether
the initial choice could influence the dynamics strongly. For such
questions, the parametrization by moments, rather than wave functions,
is important because it gives full access to the state space. For
instance, one could use a random number generator to construct the
initial moments just by restricting to those sets that obey the
Schwartz inequalities.  This will provide a systematic control to map
the whole state space, which would not be achievable by specifying
explicit wave functions. (Of course, one could randomize the
coefficients of wave functions in some basis, but the observable
meaning of those variations would be much less clear than changes of
moments.) Probing the large parameter space of the initial state is the
main problem in this context which will benefit from further numerical
support. The results of this paper thus show that quantum cosmology
provides its own set of problems which are interesting from a
numerical perspective, whose solution will then give feedback for the
specific form of the dynamics realized.

\acknowledgments

We are grateful to G.~A.~Mena Marug\'an for useful discussions.  This
work was supported in part by NSF grant 0748336 and by the Spanish
MICINN Project No. FIS2008-06078-C03-03.  DB is funded by the Spanish
Ministry of Education through \emph{Programa Nacional de Movilidad de
Recursos Humanos} from National Programme No. I-D+i2008-2011. HHH is
supported by the grant CONACyT-CB-2008-01-101774. MJK was supported by
a fellowship of the Natural Science and Engineering Research Council
of Canada. Partial support from the grant CONACyT-NSF Strong
backreaction effects in quantum cosmology is acknowledged.

\appendix

\section{Convergence of the system with an increasing
number of moments}\label{convergence}

\begin{figure}
\includegraphics[width=0.5\textwidth]{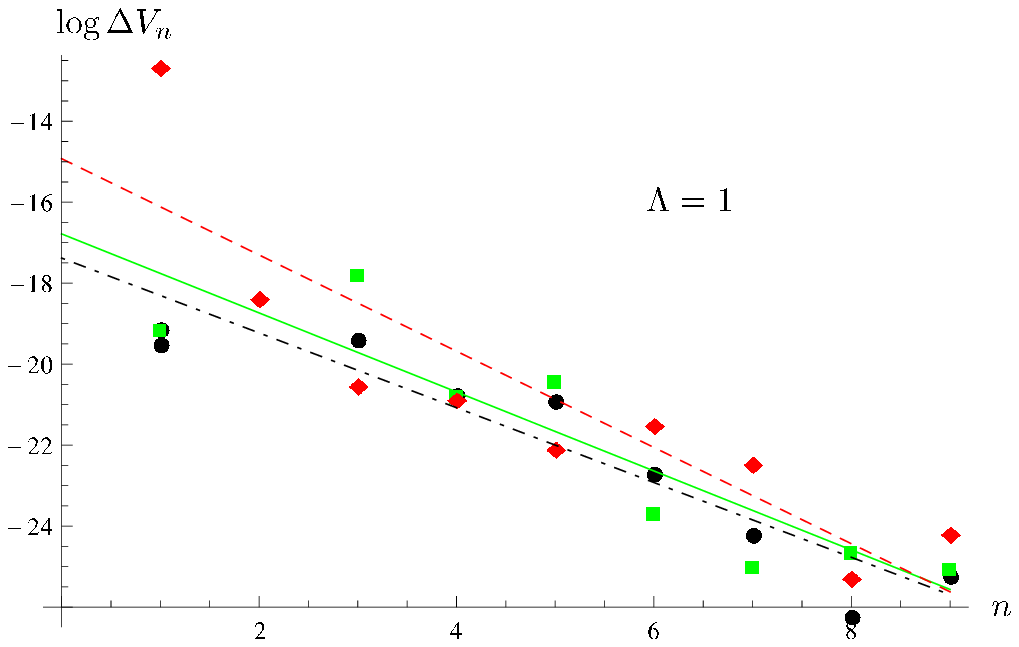}
\includegraphics[width=0.5\textwidth]{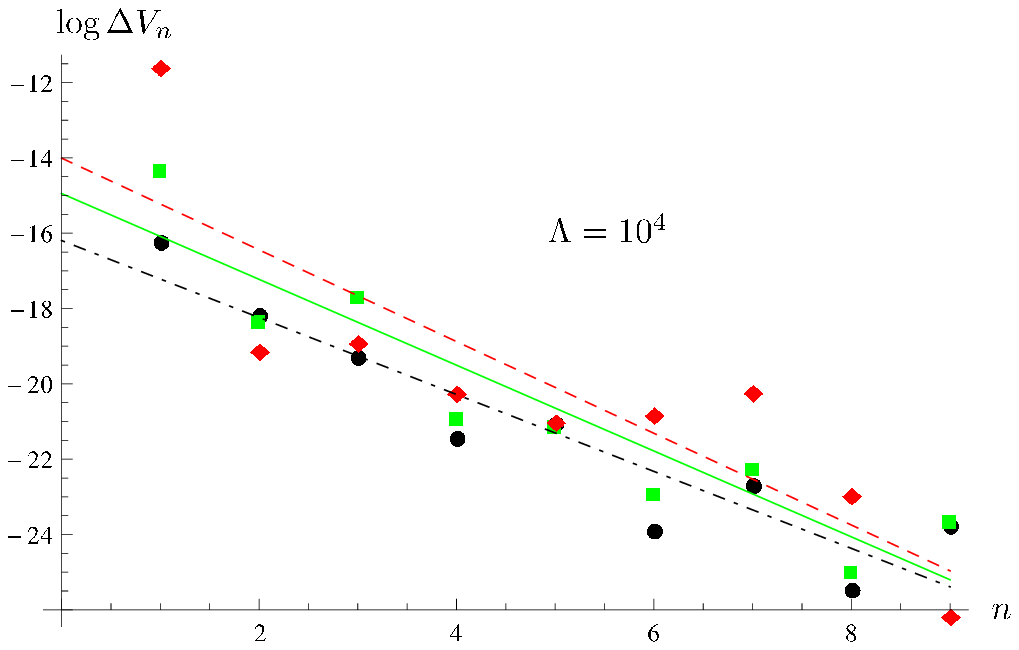}
\includegraphics[width=0.5\textwidth]{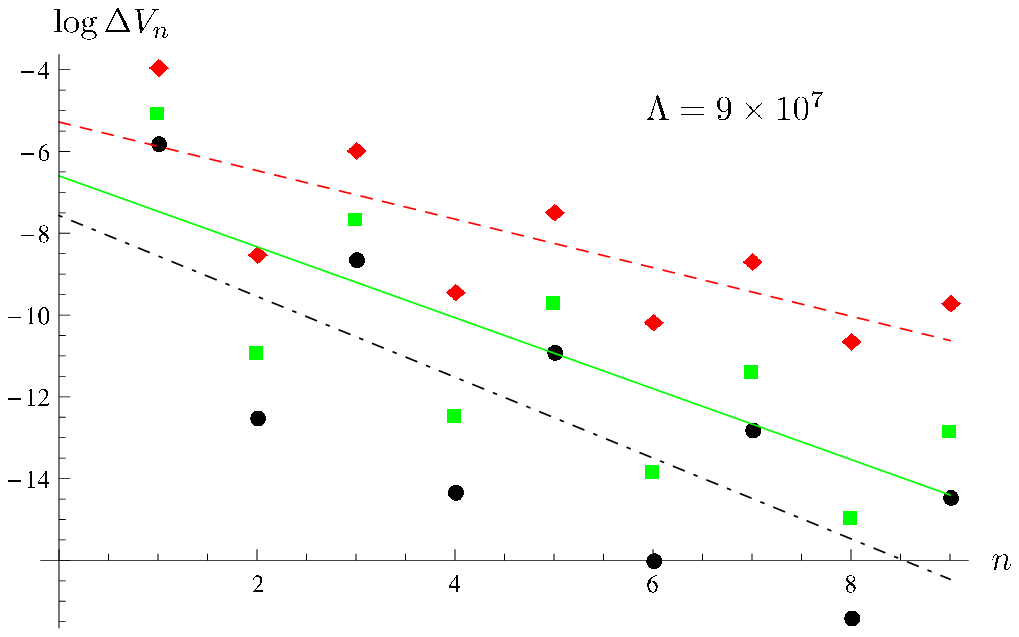}
\caption{\label{conv} In these plots we show the relative change of
the volume $\Delta V_n$ at each perturbative order for the three
different values of the cosmological constant in a logarithmic
scale. The computed results and their linear regressions at times
$\phi=0.4 \phi_{\rm fin}$ , $\phi=0.6 \phi_{\rm fin}$, and $\phi=0.9
\phi_{\rm fin}$ are plotted respectively in black (dots and dot-dashed
line), green (squares and continuous line) and red (diamonds and
dashed line).  In the first plot the points at $n=2$
corresponding to $\phi=0.4 \phi_{\rm fin}$ and $\phi=0.6 \phi_{\rm
fin}$ are missing because they are exactly (up to our numerical error)
zero. These two points have not been considered to perform the
corresponding linear regression. The slope of each linear regression
gives the convergence order. The absolute value of the different
slopes are, from upper to the lower plot and line: 1.22, 1.14, 1.02;
1.19, 0.98, 0.92; and 0.59, 0.87, 0.99.}
\end{figure}

In this appendix we address the issue of the convergence
of the solution when considering an increasing number of
moments. As we have explained in the main body of the article,
the existence of a $\hbar$ierarchy on the moments define
a semiclassical regime where the truncation of the system
into a finite number of moments makes sense. But, since we have
obtained the numerical solution of the system of equations
at different orders, another method is at our hand to check
whether the ignored moments are indeed negligible, namely
to analyze if the difference between the expectation
values at different order decreases sufficiently fast.

In order to do so, we define the relative error at
order $n$ as,
\begin{equation} \label{error}
\Delta V_n:=1- \frac{V_n}{V_{n+1}},
\end{equation}
where $V_n$ is the expectation value at order $n$. If
this object is convergent sufficiently fast with $n$, it gives
an estimate of the total relative error ($1- V_n/V_{\infty}$)
committed when truncating the system at a given order $n$.

Even though we did not write it explicitly, the object $\Delta V_n$
is time ($\phi$) dependent and, in fact, we expect it to
increase (and eventually not to converge) as we approach
the regime of large moments.
Hence we have chosen three different times to
perform the convergence tests: $0.4 \phi_{\rm fin}$,
$0.6 \phi_{\rm fin}$, and $0.9 \phi_{\rm fin}$, where
$\phi_{\rm fin}$ is the final time of each numerical
evolution and takes the value $0.893\phi_{\rm div}$ for $\Lambda=1$,
$0.807\phi_{\rm div}$ for $\Lambda=10^4$, and
$0.665\phi_{\rm div}$ for $\Lambda=9\times10^7$.

These convergence tests are shown in Fig. \ref{conv}
in a natural logarithmic scale. Firstly, we note that
the convergence is exponential in all the cases and
for all considered times. Therefore, we have performed
linear regressions for all the data.

As expected, the magnitude of the errors
increases for later times in all cases. Even so,
surprisingly for the first two plots, the convergence
is faster at those times. The numerical value of the
errors is also increasing with the value
of $\Lambda$. For instance, the relative errors of
the classical solutions ($n=1$) are of the order
$e^{-6}-e^{-4}\approx 10^{-3}-10^{-2}$ for large $\Lambda=9\times 10^7$,
whereas for small $\Lambda=1$ they are only of order
$e^{-16}-e^{-11}\approx 10^{-7}-10^{-5}$. This gives
an idea of the magnitude of the back reaction in
each case.

On the other hand, we see that our result at $10$th
order mimics very accurately the behavior of the whole quantum
system, in the sense of an asymptotic expansion. 
In particular, the largest error we find corresponds
to the large cosmological constant
case at late times ($0.9 \phi_{\rm fin}$) and
it is of the order of $e^{-9}\approx10^{-4}$.

Note that in the last plot of Fig. \ref{conv} (and also in the
second one, but less severe so) even and odd orders show different
convergence behaviors. This observation demonstrates that moments of
odd orders continue to contribute less significantly to the volume
expectation value than moments of even order, even after the state has
evolved away from Gaussian form for which odd-order moments
vanish. [For odd $n$, the expectation values in the ratio
$V_n/V_{n+1}$ in (\ref{error}) differ by even-order moments, and the
errors are seen to be enlarged in the plots.]

This analysis proves that our treatment provides a valid approximation
at all considered times, a result which strengthens the motivation to
study this general approach.

\section{High-order equations}

\begin{figure}[ht]
\includegraphics[width=0.45\textwidth]{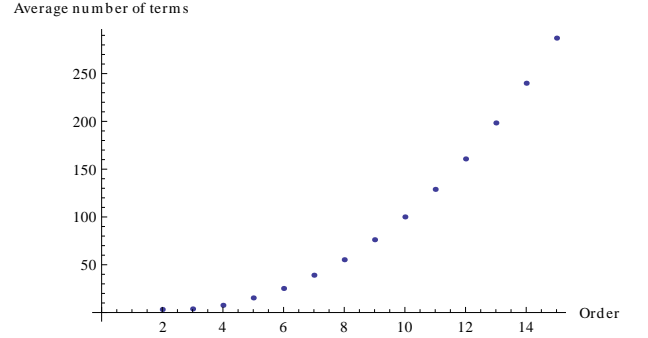}
\caption{\label{numberofterms}In this plot we show the average
number of terms that appear in each evolution equation with respect
to the order. The dependence is exponential.}
\end{figure}

In order to give a flavor of the increase in complexity of the
equations at high orders, in Fig. \ref{numberofterms} we show
the average number of terms per equation at each order. In addition,
in this appendix,
we also present explicitly the complete fifth-order equations
of motion (with $\hbar=1$).
\begin{widetext}
\begin{eqnarray*}
\dot P&=&\frac{3 \Lambda  G^{2,0}}{4 \left(P^2-\Lambda
   \right)^{3/2}}-\frac{3 P \Lambda  G^{3,0}}{4 \left(P^2-\Lambda
   \right)^{5/2}}+\frac{3 \Lambda  \left(4 P^2+\Lambda \right)
   G^{4,0}}{16 \left(P^2-\Lambda \right)^{7/2}}-\frac{3 P \Lambda 
   \left(4 P^2+3 \Lambda \right) G^{5,0}}{16 \left(P^2-\Lambda
   \right)^{9/2}}-\frac{3 \sqrt{P^2-\Lambda
   }}{2}
\\
\\ \dot V&=&\frac{9 P \Lambda  G^{2,0} V}{4
   \left(P^2-\Lambda \right)^{5/2}}-\frac{3 \Lambda  \left(4
   P^2+\Lambda \right) G^{3,0} V}{4 \left(P^2-\Lambda
   \right)^{7/2}}+\frac{15 P \Lambda  \left(4 P^2+3 \Lambda \right)
   G^{4,0} V}{16 \left(P^2-\Lambda \right)^{9/2}}
-\frac{9 \Lambda 
   \left(8 P^4+12 \Lambda  P^2+\Lambda ^2\right) G^{5,0} V}{16
   \left(P^2-\Lambda \right)^{11/2}}
\\&+&\frac{3 P V}{2
   \sqrt{P^2-\Lambda }}-\frac{3 \Lambda  G^{1,1}}{2
   \left(P^2-\Lambda \right)^{3/2}}+\frac{9 P \Lambda  G^{2,1}}{4
   \left(P^2-\Lambda \right)^{5/2}}-\frac{3 \Lambda  \left(4
   P^2+\Lambda \right) G^{3,1}}{4 \left(P^2-\Lambda
   \right)^{7/2}}+\frac{15 P \Lambda  \left(4 P^2+3 \Lambda \right)
   G^{4,1}}{16 \left(P^2-\Lambda
   \right)^{9/2}}\\ \\ \dot G^{0,2}&=&\frac{3 P
   G^{0,2}}{\sqrt{P^2-\Lambda }}-\frac{3 V \Lambda 
   G^{1,1}}{\left(P^2-\Lambda \right)^{3/2}}-\frac{3 \Lambda 
   G^{1,2}}{\left(P^2-\Lambda \right)^{3/2}}+\frac{9 P V \Lambda 
   G^{2,1}}{2 \left(P^2-\Lambda \right)^{5/2}}+\frac{9 P \Lambda 
   G^{2,2}}{2 \left(P^2-\Lambda \right)^{5/2}}-\frac{3 V \Lambda 
   \left(4 P^2+\Lambda \right) G^{3,1}}{2 \left(P^2-\Lambda
   \right)^{7/2}}
\\&-&\frac{3 \Lambda  \left(4 P^2+\Lambda \right)
   G^{3,2}}{2 \left(P^2-\Lambda \right)^{7/2}}+\frac{15 P V \Lambda 
   \left(4 P^2+3 \Lambda \right) G^{4,1}}{8 \left(P^2-\Lambda
   \right)^{9/2}}\\ \\ \dot G^{0,3}&=&-\frac{9 V G^{1,2} \Lambda }{2
   \left(P^2-\Lambda \right)^{3/2}}-\frac{9 G^{1,3} \Lambda }{2
   \left(P^2-\Lambda \right)^{3/2}}-\frac{45 P V \left(4 P^2+3
   \Lambda \right) G^{2,0} \Lambda }{16 \left(P^2-\Lambda
   \right)^{9/2}}+\frac{27 P V G^{2,2} \Lambda }{4 \left(P^2-\Lambda
   \right)^{5/2}}+\frac{27 P G^{2,3} \Lambda }{4 \left(P^2-\Lambda
   \right)^{5/2}}
\\&-&\frac{9 V \left(4 P^2+\Lambda \right) G^{3,2}
   \Lambda }{4 \left(P^2-\Lambda \right)^{7/2}}-\frac{9 P V \Lambda
   }{8 \left(P^2-\Lambda \right)^{5/2}}+\frac{9 P G^{0,3}}{2
   \sqrt{P^2-\Lambda }}+\left(\frac{9 \Lambda  \left(4 P^2+\Lambda
   \right)}{8 \left(P^2-\Lambda \right)^{7/2}}+\frac{9 \Lambda 
   G^{0,2}}{2 \left(P^2-\Lambda \right)^{3/2}}\right) G^{1,1}
\\&+&G^{0,2}
   \left(-\frac{27 P V \Lambda  G^{2,0}}{4 \left(P^2-\Lambda
   \right)^{5/2}}-\frac{27 P \Lambda  G^{2,1}}{4 \left(P^2-\Lambda
   \right)^{5/2}}+\frac{9 V \Lambda  \left(4 P^2+\Lambda \right)
   G^{3,0}}{4 \left(P^2-\Lambda \right)^{7/2}}+\frac{9 \Lambda 
   \left(4 P^2+\Lambda \right) G^{3,1}}{4 \left(P^2-\Lambda
   \right)^{7/2}}-\frac{45 P V \Lambda  \left(4 P^2+3 \Lambda
   \right) G^{4,0}}{16 \left(P^2-\Lambda
   \right)^{9/2}}\right)\\ \\ \dot G^{0,4}&=&-\frac{9 P \Lambda 
   G^{0,2}}{2 \left(P^2-\Lambda \right)^{5/2}}+\frac{6 P
   G^{0,4}}{\sqrt{P^2-\Lambda }}+\left(\frac{9 V \Lambda  \left(4
   P^2+\Lambda \right)}{2 \left(P^2-\Lambda \right)^{7/2}}+\frac{6
   \Lambda  G^{0,3}}{\left(P^2-\Lambda \right)^{3/2}}\right)
   G^{1,1}+\frac{9 \Lambda  \left(4 P^2+\Lambda \right) G^{1,2}}{2
   \left(P^2-\Lambda \right)^{7/2}}-\frac{6 V \Lambda 
   G^{1,3}}{\left(P^2-\Lambda \right)^{3/2}}\\&-&\frac{6 \Lambda 
   G^{1,4}}{\left(P^2-\Lambda \right)^{3/2}}-\frac{45 P V \Lambda 
   \left(4 P^2+3 \Lambda \right) G^{2,1}}{4 \left(P^2-\Lambda
   \right)^{9/2}}+\frac{9 P V \Lambda  G^{2,3}}{\left(P^2-\Lambda
   \right)^{5/2}}+G^{0,3} \left(-\frac{9 P V \Lambda 
   G^{2,0}}{\left(P^2-\Lambda \right)^{5/2}}-\frac{9 P \Lambda 
   G^{2,1}}{\left(P^2-\Lambda \right)^{5/2}}
\right.\\&+&\left.\frac{3 V \Lambda 
   \left(4 P^2+\Lambda \right) G^{3,0}}{\left(P^2-\Lambda
   \right)^{7/2}}+\frac{3 \Lambda  \left(4 P^2+\Lambda \right)
   G^{3,1}}{\left(P^2-\Lambda \right)^{7/2}}-\frac{15 P V \Lambda 
   \left(4 P^2+3 \Lambda \right) G^{4,0}}{4 \left(P^2-\Lambda
   \right)^{9/2}}\right)\\ \\ \dot G^{0,5}&=&\frac{45 P V \Lambda 
   \left(4 P^2+3 \Lambda \right)}{32 \left(P^2-\Lambda
   \right)^{9/2}}-\frac{225 P V \Lambda  G^{2,2} \left(4 P^2+3
   \Lambda \right)}{8 \left(P^2-\Lambda \right)^{9/2}}-\frac{45 P V
   \Lambda  G^{0,2}}{4 \left(P^2-\Lambda \right)^{5/2}}-\frac{45 P
   \Lambda  G^{0,3}}{4 \left(P^2-\Lambda \right)^{5/2}}+\frac{15 P
   G^{0,5}}{2 \sqrt{P^2-\Lambda }}
\\&+&\frac{45 V \Lambda  \left(4
   P^2+\Lambda \right) G^{1,2}}{4 \left(P^2-\Lambda
   \right)^{7/2}}+\frac{45 \Lambda  \left(4 P^2+\Lambda \right)
   G^{1,3}}{4 \left(P^2-\Lambda \right)^{7/2}}-\frac{15 V \Lambda 
   G^{1,4}}{2 \left(P^2-\Lambda \right)^{3/2}}+G^{0,4} \left(\frac{15
   \Lambda  G^{1,1}}{2 \left(P^2-\Lambda \right)^{3/2}}-\frac{45 P V
   \Lambda  G^{2,0}}{4 \left(P^2-\Lambda \right)^{5/2}}
\right.\\&-&\left.\frac{45 P
   \Lambda  G^{2,1}}{4 \left(P^2-\Lambda \right)^{5/2}}+\frac{15 V
   \Lambda  \left(4 P^2+\Lambda \right) G^{3,0}}{4 \left(P^2-\Lambda
   \right)^{7/2}}+\frac{15 \Lambda  \left(4 P^2+\Lambda \right)
   G^{3,1}}{4 \left(P^2-\Lambda \right)^{7/2}}-\frac{75 P V \Lambda 
   \left(4 P^2+3 \Lambda \right) G^{4,0}}{16 \left(P^2-\Lambda
   \right)^{9/2}}\right)\\
 \dot G^{1,1}&=&-\frac{3 V \Lambda 
   G^{2,0}}{2 \left(P^2-\Lambda \right)^{3/2}}-\frac{3 \Lambda 
   G^{2,1}}{4 \left(P^2-\Lambda \right)^{3/2}}+\frac{9 P V \Lambda 
   G^{3,0}}{4 \left(P^2-\Lambda \right)^{5/2}}+\frac{3 P \Lambda 
   G^{3,1}}{2 \left(P^2-\Lambda \right)^{5/2}}-\frac{3 V \Lambda 
   \left(4 P^2+\Lambda \right) G^{4,0}}{4 \left(P^2-\Lambda
   \right)^{7/2}}-\frac{9 \Lambda  \left(4 P^2+\Lambda \right)
   G^{4,1}}{16 \left(P^2-\Lambda \right)^{7/2}}\\&+&\frac{15 P V \Lambda
    \left(4 P^2+3 \Lambda \right) G^{5,0}}{16 \left(P^2-\Lambda
   \right)^{9/2}}\\ \\ \dot G^{1,2}&=&\frac{3 \Lambda 
   (G^{1,1})^2}{\left(P^2-\Lambda \right)^{3/2}}+\left(-\frac{9 P V
   \Lambda  G^{2,0}}{2 \left(P^2-\Lambda \right)^{5/2}}-\frac{9 P
   \Lambda  G^{2,1}}{2 \left(P^2-\Lambda \right)^{5/2}}+\frac{3 V
   \Lambda  \left(4 P^2+\Lambda \right) G^{3,0}}{2 \left(P^2-\Lambda
   \right)^{7/2}}+\frac{3 \Lambda  \left(4 P^2+\Lambda \right)
   G^{3,1}}{2 \left(P^2-\Lambda \right)^{7/2}}\right.\\&-&\left.\frac{15 P V \Lambda 
   \left(4 P^2+3 \Lambda \right) G^{4,0}}{8 \left(P^2-\Lambda
   \right)^{9/2}}\right) G^{1,1}-\frac{3 \Lambda }{8
   \left(P^2-\Lambda \right)^{3/2}}+\frac{3 P G^{1,2}}{2
   \sqrt{P^2-\Lambda }}+\left(-\frac{9 \Lambda  \left(4 P^2+\Lambda
   \right)}{16 \left(P^2-\Lambda \right)^{7/2}}-\frac{3 \Lambda 
   G^{0,2}}{4 \left(P^2-\Lambda \right)^{3/2}}\right) G^{2,0}\\&-&\frac{3
   V \Lambda  G^{2,1}}{\left(P^2-\Lambda \right)^{3/2}}-\frac{9
   \Lambda  G^{2,2}}{4 \left(P^2-\Lambda \right)^{3/2}}+\frac{9 P V
   \Lambda  G^{3,1}}{2 \left(P^2-\Lambda \right)^{5/2}}+\frac{15 P
   \Lambda  G^{3,2}}{4 \left(P^2-\Lambda \right)^{5/2}}+G^{0,2}
   \left(\frac{3 P \Lambda  G^{3,0}}{4 \left(P^2-\Lambda
   \right)^{5/2}}-\frac{3 \Lambda  \left(4 P^2+\Lambda \right)
   G^{4,0}}{16 \left(P^2-\Lambda \right)^{7/2}}\right)\\&-&\frac{3 V
   \Lambda  \left(4 P^2+\Lambda \right) G^{4,1}}{2 \left(P^2-\Lambda
   \right)^{7/2}}\\ \\ \dot G^{1,3}&=&G^{1,1} \left(\frac{9 G^{1,2}
   \Lambda }{2 \left(P^2-\Lambda \right)^{3/2}}+\frac{9 P \Lambda
   }{4 \left(P^2-\Lambda \right)^{5/2}}\right)+\frac{3 P
   G^{1,3}}{\sqrt{P^2-\Lambda }}+\left(\frac{9 V \Lambda  \left(4
   P^2+\Lambda \right)}{8 \left(P^2-\Lambda \right)^{7/2}}-\frac{3
   \Lambda  G^{0,3}}{4 \left(P^2-\Lambda \right)^{3/2}}\right)
   G^{2,0}\\&-&\frac{9 \Lambda  \left(4 P^2+\Lambda \right) G^{2,1}}{16
   \left(P^2-\Lambda \right)^{7/2}}-\frac{9 V \Lambda  G^{2,2}}{2
   \left(P^2-\Lambda \right)^{3/2}}-\frac{15 \Lambda  G^{2,3}}{4
   \left(P^2-\Lambda \right)^{3/2}}-\frac{45 P V \Lambda  \left(4
   P^2+3 \Lambda \right) G^{3,0}}{16 \left(P^2-\Lambda
   \right)^{9/2}}+\frac{27 P V \Lambda  G^{3,2}}{4 \left(P^2-\Lambda
   \right)^{5/2}}\\&+&G^{0,3} \left(\frac{3 P \Lambda  G^{3,0}}{4
   \left(P^2-\Lambda \right)^{5/2}}-\frac{3 \Lambda  \left(4
   P^2+\Lambda \right) G^{4,0}}{16 \left(P^2-\Lambda
   \right)^{7/2}}\right)+G^{1,2} \left(-\frac{27 P V \Lambda 
   G^{2,0}}{4 \left(P^2-\Lambda \right)^{5/2}}-\frac{27 P \Lambda 
   G^{2,1}}{4 \left(P^2-\Lambda \right)^{5/2}}+\frac{9 V \Lambda 
   \left(4 P^2+\Lambda \right) G^{3,0}}{4 \left(P^2-\Lambda
   \right)^{7/2}}\right.\\&+&\left.\frac{9 \Lambda  \left(4 P^2+\Lambda \right)
   G^{3,1}}{4 \left(P^2-\Lambda \right)^{7/2}}-\frac{45 P V \Lambda 
   \left(4 P^2+3 \Lambda \right) G^{4,0}}{16 \left(P^2-\Lambda
   \right)^{9/2}}\right)\\ \\ \dot G^{1,4}&=&\frac{9 \Lambda 
   \left(4 P^2+\Lambda \right)}{32 \left(P^2-\Lambda
   \right)^{7/2}}+\frac{9 V \Lambda  G^{2,1} \left(4 P^2+\Lambda
   \right)}{2 \left(P^2-\Lambda \right)^{7/2}}+\frac{9 \Lambda 
   G^{2,2} \left(4 P^2+\Lambda \right)}{8 \left(P^2-\Lambda
   \right)^{7/2}}-\frac{9 \Lambda  G^{0,2}}{4 \left(P^2-\Lambda
   \right)^{3/2}}+\frac{9 P \Lambda  G^{1,2}}{4 \left(P^2-\Lambda
   \right)^{5/2}}\\&+&G^{1,1} \left(\frac{6 \Lambda 
   G^{1,3}}{\left(P^2-\Lambda \right)^{3/2}}-\frac{9 P V \Lambda }{2
   \left(P^2-\Lambda \right)^{5/2}}\right)+\frac{9 P G^{1,4}}{2
   \sqrt{P^2-\Lambda }}-\frac{6 V \Lambda 
   G^{2,3}}{\left(P^2-\Lambda \right)^{3/2}}-\frac{45 P V \Lambda 
   \left(4 P^2+3 \Lambda \right) G^{3,1}}{4 \left(P^2-\Lambda
   \right)^{9/2}}\\&+&G^{0,4} \left(-\frac{3 \Lambda  G^{2,0}}{4
   \left(P^2-\Lambda \right)^{3/2}}+\frac{3 P \Lambda  G^{3,0}}{4
   \left(P^2-\Lambda \right)^{5/2}}-\frac{3 \Lambda  \left(4
   P^2+\Lambda \right) G^{4,0}}{16 \left(P^2-\Lambda
   \right)^{7/2}}\right)+G^{1,3} \left(-\frac{9 P V \Lambda 
   G^{2,0}}{\left(P^2-\Lambda \right)^{5/2}}-\frac{9 P \Lambda 
   G^{2,1}}{\left(P^2-\Lambda \right)^{5/2}}\right.\\&+&\left.\frac{3 V \Lambda 
   \left(4 P^2+\Lambda \right) G^{3,0}}{\left(P^2-\Lambda
   \right)^{7/2}}+\frac{3 \Lambda  \left(4 P^2+\Lambda \right)
   G^{3,1}}{\left(P^2-\Lambda \right)^{7/2}}-\frac{15 P V \Lambda 
   \left(4 P^2+3 \Lambda \right) G^{4,0}}{4 \left(P^2-\Lambda
   \right)^{9/2}}\right)\\
\dot G^{2,0}&=&-\frac{3 P
   G^{2,0}}{\sqrt{P^2-\Lambda }}+\frac{3 \Lambda  G^{3,0}}{2
   \left(P^2-\Lambda \right)^{3/2}}-\frac{3 P \Lambda  G^{4,0}}{2
   \left(P^2-\Lambda \right)^{5/2}}+\frac{3 \Lambda  \left(4
   P^2+\Lambda \right) G^{5,0}}{8 \left(P^2-\Lambda
   \right)^{7/2}}\\
\dot G^{2,1}&=&\left(-\frac{9 P
   \Lambda  G^{2,1}}{4 \left(P^2-\Lambda \right)^{5/2}}+\frac{3 V
   \Lambda  \left(4 P^2+\Lambda \right) G^{3,0}}{4 \left(P^2-\Lambda
   \right)^{7/2}}+\frac{3 \Lambda  \left(4 P^2+\Lambda \right)
   G^{3,1}}{4 \left(P^2-\Lambda \right)^{7/2}}-\frac{15 P V \Lambda 
   \left(4 P^2+3 \Lambda \right) G^{4,0}}{16 \left(P^2-\Lambda
   \right)^{9/2}}\right) G^{2,0}\\&-&\frac{9 P V \Lambda 
   (G^{2,0})^2}{4 \left(P^2-\Lambda \right)^{5/2}}-\frac{3 P G^{2,1}}{2
   \sqrt{P^2-\Lambda }}+\left(\frac{3 P \Lambda  G^{1,1}}{2
   \left(P^2-\Lambda \right)^{5/2}}-\frac{3 V \Lambda }{2
   \left(P^2-\Lambda \right)^{3/2}}\right) G^{3,0}+\frac{9 P V
   \Lambda  G^{4,0}}{4 \left(P^2-\Lambda \right)^{5/2}}\\&-&\frac{3
   \Lambda  \left(4 P^2+\Lambda \right) G^{1,1} G^{4,0}}{8
   \left(P^2-\Lambda \right)^{7/2}}+\frac{3 P \Lambda  G^{4,1}}{4
   \left(P^2-\Lambda \right)^{5/2}}-\frac{3 V \Lambda  \left(4
   P^2+\Lambda \right) G^{5,0}}{4 \left(P^2-\Lambda
   \right)^{7/2}}\\
\dot G^{2,2}&=&-\frac{9 P \Lambda 
   (G^{2,1})^2}{2 \left(P^2-\Lambda \right)^{5/2}}+\frac{3 \Lambda 
   G^{1,1} G^{2,1}}{\left(P^2-\Lambda \right)^{3/2}}+\left(\frac{3 V
   \Lambda  \left(4 P^2+\Lambda \right) G^{3,0}}{2 \left(P^2-\Lambda
   \right)^{7/2}}+\frac{3 \Lambda  \left(4 P^2+\Lambda \right)
   G^{3,1}}{2 \left(P^2-\Lambda \right)^{7/2}}-\frac{15 P V \Lambda 
   \left(4 P^2+3 \Lambda \right) G^{4,0}}{8 \left(P^2-\Lambda
   \right)^{9/2}}\right) G^{2,1}\\&+&G^{2,0} \left(-\frac{3 G^{1,2}
   \Lambda }{2 \left(P^2-\Lambda \right)^{3/2}}-\frac{9 P V G^{2,1}
   \Lambda }{2 \left(P^2-\Lambda \right)^{5/2}}+\frac{9 P \Lambda
   }{4 \left(P^2-\Lambda \right)^{5/2}}\right)-\frac{9 \Lambda 
   \left(4 P^2+\Lambda \right) G^{3,0}}{8 \left(P^2-\Lambda
   \right)^{7/2}}-\frac{3 V \Lambda  G^{3,1}}{\left(P^2-\Lambda
   \right)^{3/2}}-\frac{3 \Lambda  G^{3,2}}{2 \left(P^2-\Lambda
   \right)^{3/2}}\\&+&G^{1,2} \left(\frac{3 P \Lambda  G^{3,0}}{2
   \left(P^2-\Lambda \right)^{5/2}}-\frac{3 \Lambda  \left(4
   P^2+\Lambda \right) G^{4,0}}{8 \left(P^2-\Lambda
   \right)^{7/2}}\right)+\frac{9 P V \Lambda  G^{4,1}}{2
   \left(P^2-\Lambda \right)^{5/2}}\\ \\ \dot G^{2,3}&=&G^{2,1}
   \left(\frac{45 P \Lambda }{8 \left(P^2-\Lambda
   \right)^{5/2}}-\frac{27 P \Lambda  G^{2,2}}{4 \left(P^2-\Lambda
   \right)^{5/2}}\right)+G^{2,0} \left(-\frac{3 G^{1,3} \Lambda }{2
   \left(P^2-\Lambda \right)^{3/2}}-\frac{27 P V G^{2,2} \Lambda }{4
   \left(P^2-\Lambda \right)^{5/2}}-\frac{9 P V \Lambda }{8
   \left(P^2-\Lambda \right)^{5/2}}\right)\\&+&G^{1,1} \left(\frac{9
   \Lambda  G^{2,2}}{2 \left(P^2-\Lambda \right)^{3/2}}-\frac{9
   \Lambda }{4 \left(P^2-\Lambda \right)^{3/2}}\right)+\frac{3 P
   G^{2,3}}{2 \sqrt{P^2-\Lambda }}+\frac{9 V \Lambda  \left(4
   P^2+\Lambda \right) G^{3,0}}{8 \left(P^2-\Lambda
   \right)^{7/2}}-\frac{9 \Lambda  \left(4 P^2+\Lambda \right)
   G^{3,1}}{4 \left(P^2-\Lambda \right)^{7/2}}\\&-&\frac{9 V \Lambda 
   G^{3,2}}{2 \left(P^2-\Lambda \right)^{3/2}}-\frac{45 P V \Lambda 
   \left(4 P^2+3 \Lambda \right) G^{4,0}}{16 \left(P^2-\Lambda
   \right)^{9/2}}+G^{1,3} \left(\frac{3 P \Lambda  G^{3,0}}{2
   \left(P^2-\Lambda \right)^{5/2}}-\frac{3 \Lambda  \left(4
   P^2+\Lambda \right) G^{4,0}}{8 \left(P^2-\Lambda
   \right)^{7/2}}\right)\\&+&G^{2,2} \left(\frac{9 V \Lambda  \left(4
   P^2+\Lambda \right) G^{3,0}}{4 \left(P^2-\Lambda
   \right)^{7/2}}+\frac{9 \Lambda  \left(4 P^2+\Lambda \right)
   G^{3,1}}{4 \left(P^2-\Lambda \right)^{7/2}}-\frac{45 P V \Lambda 
   \left(4 P^2+3 \Lambda \right) G^{4,0}}{16 \left(P^2-\Lambda
   \right)^{9/2}}\right)\\
\dot G^{3,0}&=&-\frac{9 \Lambda 
   (G^{2,0})^2}{4 \left(P^2-\Lambda \right)^{3/2}}+\left(\frac{9 P
   \Lambda  G^{3,0}}{4 \left(P^2-\Lambda \right)^{5/2}}-\frac{9
   \Lambda  \left(4 P^2+\Lambda \right) G^{4,0}}{16
   \left(P^2-\Lambda \right)^{7/2}}\right) G^{2,0}-\frac{9 P
   G^{3,0}}{2 \sqrt{P^2-\Lambda }}+\frac{9 \Lambda  G^{4,0}}{4
   \left(P^2-\Lambda \right)^{3/2}}-\frac{9 P \Lambda  G^{5,0}}{4
   \left(P^2-\Lambda \right)^{5/2}}\\ \\ \dot G^{3,1}&=&\frac{3 V
   \Lambda  \left(4 P^2+\Lambda \right) (G^{3,0})^2}{4
   \left(P^2-\Lambda \right)^{7/2}}+\frac{3 \Lambda  G^{1,1}
   G^{3,0}}{2 \left(P^2-\Lambda \right)^{3/2}}+\left(\frac{3 \Lambda
    \left(4 P^2+\Lambda \right) G^{3,1}}{4 \left(P^2-\Lambda
   \right)^{7/2}}-\frac{15 P V \Lambda  \left(4 P^2+3 \Lambda
   \right) G^{4,0}}{16 \left(P^2-\Lambda \right)^{9/2}}\right)
   G^{3,0}\\&+&G^{2,0} \left(-\frac{9 \Lambda  G^{2,1}}{4
   \left(P^2-\Lambda \right)^{3/2}}-\frac{9 P V \Lambda  G^{3,0}}{4
   \left(P^2-\Lambda \right)^{5/2}}\right)-\frac{3 P
   G^{3,1}}{\sqrt{P^2-\Lambda }}-\frac{3 V \Lambda  G^{4,0}}{2
   \left(P^2-\Lambda \right)^{3/2}}-\frac{9 \Lambda  \left(4
   P^2+\Lambda \right) G^{2,1} G^{4,0}}{16 \left(P^2-\Lambda
   \right)^{7/2}}\\&+&\frac{3 \Lambda  G^{4,1}}{4 \left(P^2-\Lambda
   \right)^{3/2}}+\frac{9 P V \Lambda  G^{5,0}}{4 \left(P^2-\Lambda
   \right)^{5/2}}\\ \\ \dot G^{3,2}&=&\frac{3 \Lambda  \left(4
   P^2+\Lambda \right) (G^{3,1})^2}{2 \left(P^2-\Lambda
   \right)^{7/2}}+\frac{3 \Lambda  G^{1,1} G^{3,1}}{\left(P^2-\Lambda
   \right)^{3/2}}+\left(-\frac{9 P \Lambda  G^{2,1}}{2
   \left(P^2-\Lambda \right)^{5/2}}-\frac{15 P V \Lambda  \left(4
   P^2+3 \Lambda \right) G^{4,0}}{8 \left(P^2-\Lambda
   \right)^{9/2}}\right) G^{3,1}\\&+&G^{2,0} \left(-\frac{9 G^{2,2}
   \Lambda }{4 \left(P^2-\Lambda \right)^{3/2}}-\frac{9 P V G^{3,1}
   \Lambda }{2 \left(P^2-\Lambda \right)^{5/2}}-\frac{9 \Lambda }{8
   \left(P^2-\Lambda \right)^{3/2}}\right)+G^{3,0} \left(\frac{3 V
   \left(4 P^2+\Lambda \right) G^{3,1} \Lambda }{2 \left(P^2-\Lambda
   \right)^{7/2}}+\frac{27 P \Lambda }{8 \left(P^2-\Lambda
   \right)^{5/2}}\right)\\&-&\frac{3 P G^{3,2}}{2 \sqrt{P^2-\Lambda
   }}-\frac{27 \Lambda  \left(4 P^2+\Lambda \right) G^{4,0}}{16
   \left(P^2-\Lambda \right)^{7/2}}+G^{2,2} \left(\frac{9 P \Lambda 
   G^{3,0}}{4 \left(P^2-\Lambda \right)^{5/2}}-\frac{9 \Lambda 
   \left(4 P^2+\Lambda \right) G^{4,0}}{16 \left(P^2-\Lambda
   \right)^{7/2}}\right)-\frac{3 V \Lambda 
   G^{4,1}}{\left(P^2-\Lambda
   \right)^{3/2}}\\ \\ \dot G^{4,0}&=&\frac{3 P \Lambda 
   (G^{3,0})^2}{\left(P^2-\Lambda \right)^{5/2}}-\frac{3 \Lambda 
   G^{2,0} G^{3,0}}{\left(P^2-\Lambda \right)^{3/2}}-\frac{3 \Lambda 
   \left(4 P^2+\Lambda \right) G^{4,0} G^{3,0}}{4 \left(P^2-\Lambda
   \right)^{7/2}}-\frac{6 P G^{4,0}}{\sqrt{P^2-\Lambda }}+\frac{3
   \Lambda  G^{5,0}}{\left(P^2-\Lambda
   \right)^{3/2}}\\ \\ \dot G^{4,1}&=&-\frac{15 P V \Lambda  \left(4
   P^2+3 \Lambda \right) (G^{4,0})^2}{16 \left(P^2-\Lambda
   \right)^{9/2}}+\frac{3 \Lambda  G^{1,1} G^{4,0}}{2
   \left(P^2-\Lambda \right)^{3/2}}-\frac{9 P \Lambda  G^{2,1}
   G^{4,0}}{4 \left(P^2-\Lambda \right)^{5/2}}+\frac{3 V \Lambda 
   \left(4 P^2+\Lambda \right) G^{3,0} G^{4,0}}{4 \left(P^2-\Lambda
   \right)^{7/2}}+\frac{3 P \Lambda  G^{3,0}
   G^{3,1}}{\left(P^2-\Lambda \right)^{5/2}}\\&+&G^{2,0} \left(-\frac{3
   \Lambda  G^{3,1}}{\left(P^2-\Lambda \right)^{3/2}}-\frac{9 P V
   \Lambda  G^{4,0}}{4 \left(P^2-\Lambda
   \right)^{5/2}}\right)-\frac{9 P G^{4,1}}{2 \sqrt{P^2-\Lambda
   }}-\frac{3 V \Lambda  G^{5,0}}{2 \left(P^2-\Lambda
   \right)^{3/2}}\\ \\ \dot G^{5,0}&=&-\frac{15 \Lambda  \left(4
   P^2+\Lambda \right) (G^{4,0})^2}{16 \left(P^2-\Lambda
   \right)^{7/2}}-\frac{15 \Lambda  G^{2,0} G^{4,0}}{4
   \left(P^2-\Lambda \right)^{3/2}}+\frac{15 P \Lambda  G^{3,0}
   G^{4,0}}{4 \left(P^2-\Lambda \right)^{5/2}}-\frac{15 P G^{5,0}}{2
   \sqrt{P^2-\Lambda }}
\end{eqnarray*}
\end{widetext}

\end{document}